\documentclass[apj]{emulateapj}
\usepackage{amsmath}
\usepackage{natbib}

\newcommand{\HII}{H {\small{II}} }
\newcommand{\kms}{{\rm km~s}^{-1}}
\newcommand{\Msun} {M_\sun}

\newcommand{\mjyb}{{\rm mJy~beam}^{-1}}

\newcommand{\jyb}{{\rm Jy~beam}^{-1}}
\newcommand{\mjybkms}{{\rm mJy~beam}^{-1}{\rm km~s}^{-1}}
\newcommand{\jybkms}{{\rm Jy~beam}^{-1}{\rm km~s}^{-1}}

\newcommand{\dotsec}{\rlap.{''}}
\newcommand{\dotdeg}{\rlap.{^\circ}}

\shorttitle{Hypercompact \HII region G35.58-0.03}
\shortauthors{ZHANG ET AL.}

\begin{document}


\title{Submillimeter Array and Very Large Array Observations \\in the Hypercompact \HII region G35.58-0.03}


\author{Chuan-Peng Zhang\altaffilmark{1,2,3,4}, Jun-Jie Wang\altaffilmark{1,3}, Jin-Long Xu\altaffilmark{1,3}, Friedrich Wyrowski\altaffilmark{2}, Karl M. Menten\altaffilmark{2}}

\email{cpzhang@mpifr-bonn.mpg.de}


\altaffiltext{1}{National Astronomical Observatories, Chinese
Academy of Sciences, 100012 Beijing, China}
\altaffiltext{2}{Max-Planck-Institut f\"ur Radioastronomie, Auf dem H\"ugel 69, D-53121 Bonn, Germany}
\altaffiltext{3}{NAOC-TU Joint Center for Astrophysics, 850000
Lhasa, China}
\altaffiltext{4}{University of the Chinese Academy
of Sciences, 100080 Beijing, China}



\begin{abstract}

The formation of hypercompact (HC) \HII regions is an important stage
in massive star formation. Spectral line and continuum observations
can explore its dynamic conditions. We present high angular
resolution observations carried out with the Submillimeter Array
(SMA) and the Very Large Array (VLA) toward the HC \HII region
G35.58-0.03. With the 1.3 mm SMA and 1.3 cm VLA, we detected a total
of about 25 transitions of 8 different species and their isotopologues
(CO, CH$_3$CN, SO$_2$, CH$_3$CCH, OCS, CS, H, and NH$_{3}$).
G35.58-0.03 consists of an HC \HII core with electron temperature
$T_e^*\geqq5500$\,K, emission measure EM $\approx
1.9\times10^{9}$\,pc~cm$^{-6}$, local volume electron density $n_e=
3.3\times10^{5}$\,cm$^{-3}$, and a same width of radio recombination line
FWHM $\approx$ 43.2~$\kms$ for both H30$\alpha$ and H38$\beta$ at its intrinsic core size
$\sim$3714 AU. The H30$\alpha$ line shows evidence of an ionized outflow driving a molecular outflow.
Based on the derived Lyman continuum flux, there should 
be an early-type star equivalent to O6.5 located inside the \HII region.
From the continuum spectral energy distribution from
3.6 cm, 2.0 cm, 1.3 cm, 1.3 mm and 0.85 mm to 0.45 mm, we distinguished 
the free-free emission (25\% $\sim$ 55\%) from the warm dust component (75\% $\sim$ 45\%) at 1.3 mm. 
The molecular envelope shows
evidence of infall and outflow with an infall rate 0.05
M$_{\sun}$~yr$^{-1}$ and a mass loss rate $5.2\times10^{-3}$ ${\rm
M_{\sun}~yr^{-1}}$. The derived momentum ($\sim0.05$ M$_{\sun}\,\kms$) is consistent between the 
infalling and outflowing gas per year. It is suggested that the infall is predominant and the envelope
mass of dense core is increasing rapidly, but the accretion in the inner part might already be halted.

\end{abstract}


\section{Introduction} \label{sect:intro}

Massive star formation ($\gtrsim$ 10 $\Msun$) is difficult to
understand because of the large distance ($\gtrsim$ 1 kpc), high
extinction ($A_V$ $\gtrsim$ 100 mag), and short evolutionary
timescale ($\lesssim$ 10$^4$ yr) \citep{kurt2005,vand2005} of massive star formation regions.
High angular resolution millimeter observations are able to unveil
the physical envelope conditions of individual cores. High angular
resolution centimeter observations can penetrate the dense dust emission of cores to search for hypercompact (HC) \HII or ultracompact
(UC) \HII regions excited by the embedded protostar.

HC \HII regions are associated with the earliest stages that the central object has a mass equivalent to an O star. Early B- and O-type stars are usually found
to be deeply embedded in dense molecular clouds where they produce HC \HII or
UC \HII regions. The HC \HII regions often show roughly linear continuum spectral 
energy distributions (SEDs) with frequency $S_{\nu} \varpropto
\nu^{\alpha}$ in the region up to $\sim$100 GHz \citep{fran2000,beut2004,keto2008,galv2009}.
The variation of index $\alpha$ can trace the density
gradients in the ionized gas \citep{fran2000}. Typically HC \HII regions have  
small size ($\lesssim$ 0.03 pc), high electron
density ($10^5-10^6$ cm$^{-3}$), high emission measure ($\gtrsim
10^8$ pc cm$^{-6}$), very broad radio recombination lines (FWHM
$\gtrsim$ 40 $\kms$) \citep{kurt2000,beut2007,sewi2011,choi2012}. HC
\HII regions are usually more advanced in evolution than hot molecular cores
and younger than UC \HII regions \citep{kurt2000}. However, there are some HC \HII regions 
(presumably those where the central stars are still accreting) where 
the HC \HII region coexists with a hot core.
HC \HII regions will evolve into UC \HII regions with expanding their size.
Unveiling the evolution process of HC \HII regions is helpful to understand 
massive star formation. However, few HC
\HII regions are known to investigate the
evolutionary process of massive star formation.

\begin{figure*}
\figurenum{1}
\begin{center}
\includegraphics[angle=0,scale=0.8]{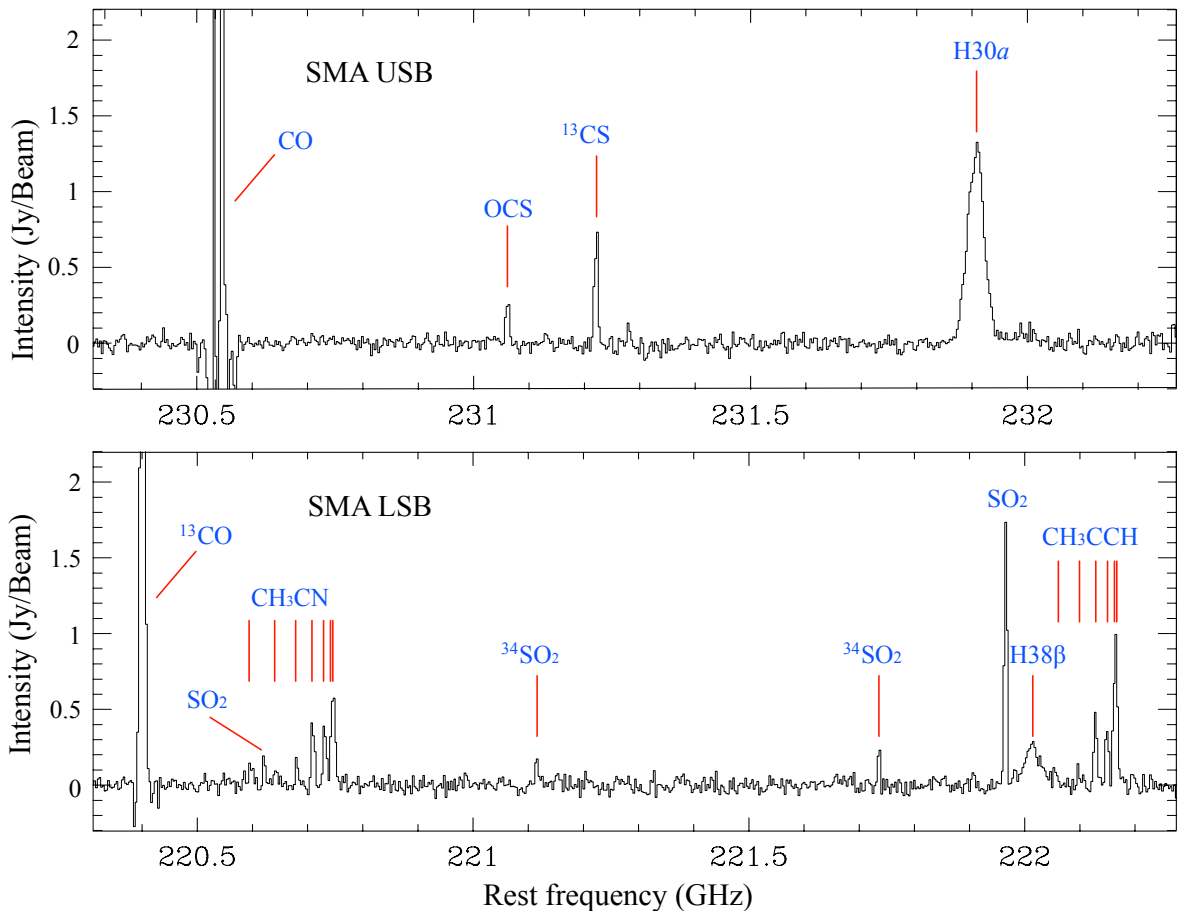}
\caption{ Wide-band SMA spectra extracted from the line data-cubes
in the image domain at the position of the 1.3 mm continuum peak.
The channel spacing in this plot is 4.25$~\kms$. }
\label{fig_usb_lsb}
\end{center}
\end{figure*}

Usually, the massive star formation process is accompanied by infall, outflow,
and/or rotation movements. Infall and accretion increase the mass of
massive stars. Outflows transfer angular momentum of infalling gas \citep{keto2002,keto2007}.
High angular resolution Submillimeter Array (SMA) and Very Large Array (VLA) observation in millimeter and centimeter wavelength will help to further understand these dynamical phenomena.

The HC \HII region G35.58-0.03 is located at the far kinematic distance
of 10.2 kpc \citep{fish2003,wats2003}. Water and OH masers
\citep{fors1989,debu2005,fish2005}, but no methanol masers
\citep{casw1995}, have been detected at this site. VLA 2 and 3.6 cm
maps indicate that G35.58-0.03 is just resolved into two extremely
close UC \HII regions: western G35.578-0.030 and eastern
G35.578-0.031 \citep{kurt1994}. The low resolution 3.6 and 21 cm maps
show that both G35.578-0.030 and G35.578-0.031 are lying within 
large-scale extended continuum
emission \citep{kurt1999,debu2005}. In addition, \citet{plum1992},
\citet{muel2002}, \citet{shir2003}, and \citet{n68} have
investigated G35.58-0.03 with low angular resolution, but they
did not present its dynamical information, even not to resolve it.

\begin{figure}
\figurenum{2}
\begin{center}
\includegraphics[angle=0,scale=0.85]{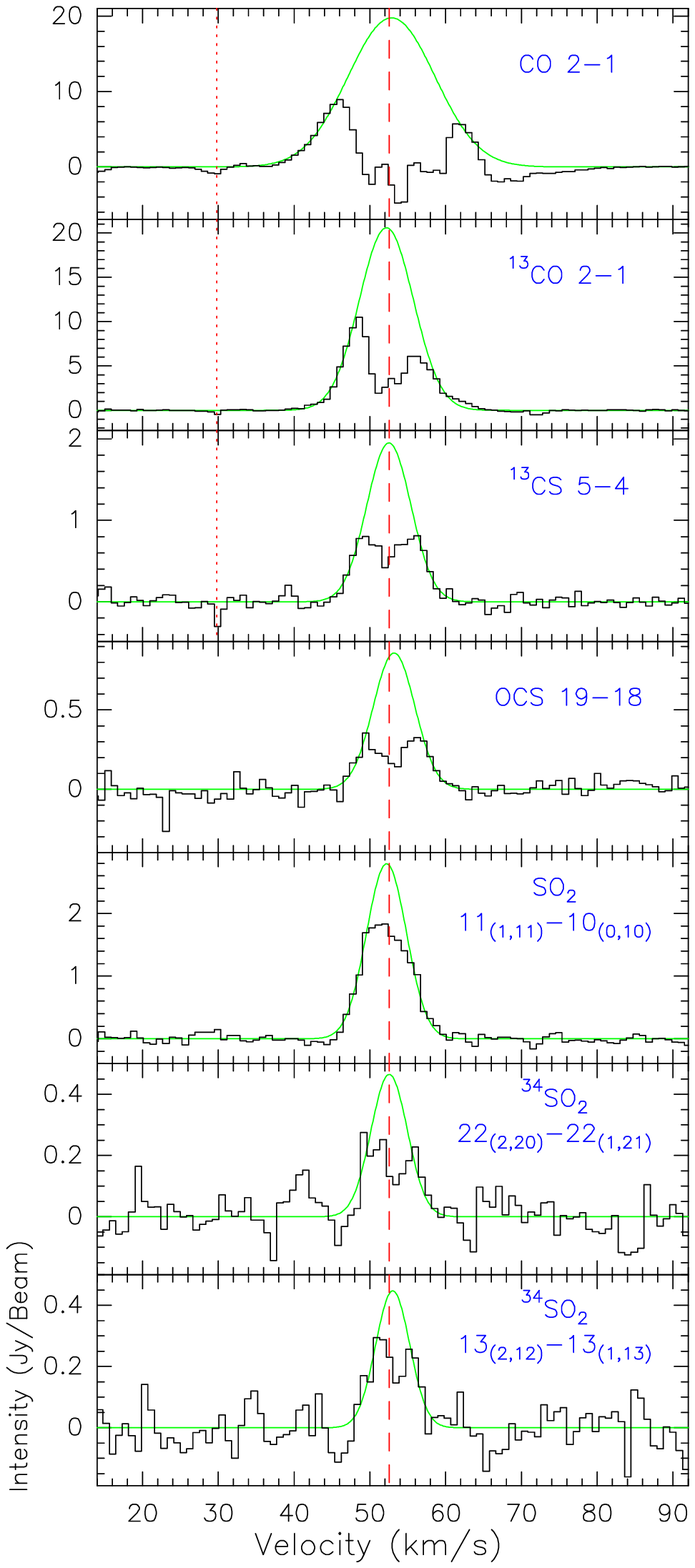}
\end{center}
\caption{ Molecular spectra at the position of the 1.3 mm continuum
peak. The channel spacing in the plots is 0.85$~\kms$. The red
dashed line crossing all the spectra denotes the systemic velocity
($V_{sys} = 52.5~\kms$), and the red dotted line indicates one
absorption dip crossing spectra CO, $^{13}$CO, and $^{13}$CS. The
green curves are Gaussian fitting lines.} \label{fig_spectra}
\end{figure}

In this work we mainly report on 1.3 millimeter (mm) and 1.3
centimeter (cm) interferometric observations performed with the SMA
and the VLA at angular resolutions of $\sim$3$\dotsec$4 and
$\sim$0$\dotsec$3, respectively, toward HC \HII region G35.58-0.03.
In Section \ref{sect:obs}, we describe the observations and reduction
of archival SMA 1.3 mm and VLA 1.3+3.6 cm data. We present the main
results from observations of spectral lines, moments maps, and
position-velocity (PV) diagrams in Section \ref{sect:result}. In Section \ref{sect:disc}, we
exhibit and discuss some results concerning the physical and dynamical 
conditions of the HC \HII region. Finally, conclusions are given in Section
\ref{sect:conclu}.

\section{Observations and Data} \label{sect:obs}

\subsection{SMA} \label{sect:sma}

\begin{center}
\begin{deluxetable*}{lcccccc}
\tabletypesize{\scriptsize} \tablecaption{Line parameters and
Gaussian fitting results\tablenotemark{a} \tablenotemark{b}} \tablewidth{0pt}
\tablehead{
\colhead{Molecule} & \colhead{Transition} & \colhead{Frequency}  &  \colhead{Flux}  &  \colhead{Velocity}   &  \colhead{FWHM} & \colhead{Intensity}  \\
                   &                      & GHz                &  Jy/beam$\cdot$km/s     &   km/s            & km/s        &   Jy/beam          \\
(1)  &   (2)   &  (3)  &  (4) &  (5) & (6) & (7) }
\startdata
NH$_3$       &   $(2,2)$                & 23.722633    &    -0.02(0.01)     &   47.44(0.58)   &   2.92(1.47)      & -0.007(0.001)    \\
NH$_3$       &   $(2,2)$                & 23.722633    &    -0.16(0.01)     &   52.60(0.17)   &   4.61(0.43)      & -0.033(0.001)    \\
NH$_3$       &   $(3,3)$                & 23.870129    &    -0.04(0.03)     &   48.17(2.26)   &   5.24(3.32)      & -0.008(0.006)     \\
NH$_3$       &   $(3,3)$                & 23.870129    &    -0.12(0.04)     &   53.17(0.35)   &   4.35(0.64)      & -0.026(0.006)     \\
$^{13}$CO    &   2-1                    & 220.39868    &    183.90(2.19)    &   52.19(0.02)   &   8.39(0.05)      & 20.60(0.37)     \\
CH$_3$CN     & 12$_6$-11$_6$            & 220.59444    &    ---             &   ---           &   ---             & ---           \\
SO$_2$       & 56$_{5,51}$-57$_{4,54}$  & 220.61850    &    18.47(0.58)     &   52.21(0.05)   &   6.21(0.17)      & 2.79(0.09)      \\
CH$_3$CN     & 12$_5$-11$_5$            & 220.64110    &    ---             &   ---           &   ---             & ---           \\
CH$_3$CN     & 12$_4$-11$_4$            & 220.67930    &    2.28(0.49)      &   122.80(0.50)  &   6.62(1.22)      & 0.32(0.06)     \\
CH$_3$CN     & 12$_3$-11$_3$            & 220.70902    &    5.99(0.44)      &   82.43(0.22)   &   8.01(0.53)      & 0.70(0.06)     \\
CH$_3$CN     & 12$_2$-11$_2$            & 220.73027    &    4.91(0.35)      &   53.41(0.24)   &   8.19(0.76)      & 0.56(0.06)      \\
CH$_3$CN     & 12$_1$-11$_1$            & 220.74302    &    ---             &   ---           &   ---             & ---            \\
CH$_3$CN     & 12$_0$-11$_0$            & 220.74727    &    ---             &   ---           &   ---             & ---             \\
$^{34}$SO$_2$& 22$_{2,20}$-22$_{1,21}$  & 221.11490    &    2.81(0.72)      &   52.56(0.38)   &   5.66(1.01)      & 0.47(0.06)      \\
$^{34}$SO$_2$& 13$_{2,12}$-13$_{1,13}$  & 221.73571    &    2.53(0.32)      &   53.02(0.23)   &   5.30(0.66)      & 0.45(0.04)      \\
SO$_2$       & 11$_{1,11}$-10$_{0,10}$  & 221.96520    &    ---             &   ---           &   ---             & ---            \\
H            & 38$\beta$                & 222.01175    &    12.11(0.69)     &   49.87(1.16)   &  43.18(3.05)      & 0.26(0.04)            \\
CH$_3$CCH    & 13$_5$-12$_5$            & 222.06103    &    ---             &   ---           &   ---             & ---            \\
CH$_3$CCH    & 13$_4$-12$_4$            & 222.09915    &    0.95(0.33)      &   123.00(0.75)  &   3.99(1.11)      & 0.22(0.08)      \\
CH$_3$CCH    & 13$_3$-12$_3$            & 222.12881    &    5.64(0.38)      &   81.93(0.11)   &   7.44(0.60)      & 0.71(0.08)      \\
CH$_3$CCH    & 13$_2$-12$_2$            & 222.15001    &    4.94(0.37)      &   53.84(0.01)   &   6.55(0.10)      & 0.71(0.08)      \\
CH$_3$CCH    & 13$_1$-12$_1$            & 222.16273    &    ---             &   ---           &   ---             & ---            \\
CH$_3$CCH    & 13$_0$-12$_0$            & 222.16697    &    ---             &   ---           &   ---             & ---            \\
CO           & 2-1                      & 230.53800    &    284.60(1.78)    &   52.86(0.02)   &   13.53(0.05)     & 19.80(0.51)     \\
OCS          & 19-18                    & 231.06099    &    5.94(1.46)      &   53.20(0.20)   &   6.49(0.75)      & 0.86(0.04)      \\
$^{13}$CS    & 5-4                      & 231.22077    &    14.27(1.28)     &   52.52(0.11)   &   6.88(0.40)      & 1.95(0.05)      \\
H            & 30$\alpha$               & 231.90090    &    61.37(0.68)     &   45.03(0.23)   &   43.16(0.42)     & 1.34(0.05)      \\
\hline
\enddata
\tablenotetext{a}{All lines are indicated in Figure
\ref{fig_usb_lsb}. The Gaussian fitting lines are exhibited in
Figures \ref{fig_spectra} and \ref{fig_nh3_spec}. }
\tablenotetext{b}{The DPFU is $\sim$2.55$\times$10$^{4}$ K per Jy beam$^{-1}$ for VLA data, and $\sim$2.22 K  per Jy beam$^{-1}$ for SMA data. }
\label{tab_lines}
\end{deluxetable*}
\end{center}

The 1.3 mm data for G35.58-0.03 are publicly 
available in the SMA data
archive\footnote{http://cfa-www.harvard.edu/rtdc/index-sma.html}
\citep{ho2004}, and were observed on 2008 June 22 in its compact 
configuration\footnote{The SMA is a joint project between the
Smithsonian Astrophysical Observatory and the Academia Sinica
Institute of Astronomy and Astrophysics and is funded by the
Smithsonian Institution and the Academia Sinica.}. Two sidebands
covered the frequency ranges of $220.3-222.3$ GHz and $230.3-232.3$
GHz with a frequency resolution of $\approx 0.812$ MHz (or velocity
resolution of $0.85~\kms$). The approximate synthesized beam size 
(full width at half-power) is $3\dotsec53 \times 3\dotsec19$ with position angle
(P.A.) = 76$\dotdeg$3 for lower sideband, and $3\dotsec38 \times 3\dotsec05$ with P.A. 
= 76$\dotdeg$1 for upper sideband. The SMA primary beam at 230 GHz is $\sim$55$''$. 
The relevant degrees per flux unit factor is DPFU$_{\rm 1.3\,mm} \sim 2.22$ K per Jy beam$^{-1}$.

The phase tracking center was $\alpha$(J2000) =
18$\mathrm{^h}$56$\mathrm{^m}$22$\rlap.{^{\mathrm{s}}}$533 and
$\delta$(J2000) = 02$^\circ$20$'$27$\dotsec$50. QSO 3C279 and Uranus
were used as bandpass and flux calibrators. QSO J1751+096 and
QSO J1830+063 were observed for the antenna gain corrections. The
calibration and imaging were performed in
Miriad\footnote{http://sma-www.cfa.harvard.edu/miriad}. 
We summed up the line-free channels, and produced a ``pseudo'' continuum 
database that was subtracted from the $uv$-database in the $uv$-plane. 
The line-free continuum was self-calibrated, and
the gain solutions were applied to the spectral line data. The
image data cubes were exported to CLASS in
GILDAS\footnote{http://www.iram.fr/IRAMFR/GILDAS/} for further
spectral line processing. The continuum-free spectra across the entire
sidebands at the position of 1.3 mm peak are shown in Figure
\ref{fig_usb_lsb}. The rms noise in the final images is in the range of 
[40, 80]$~\mjyb$ for the line data and $\sim$19$~\mjyb$ for the continuum data.

\begin{figure}
\figurenum{3}
\begin{center}
\includegraphics[angle=0,scale=0.85]{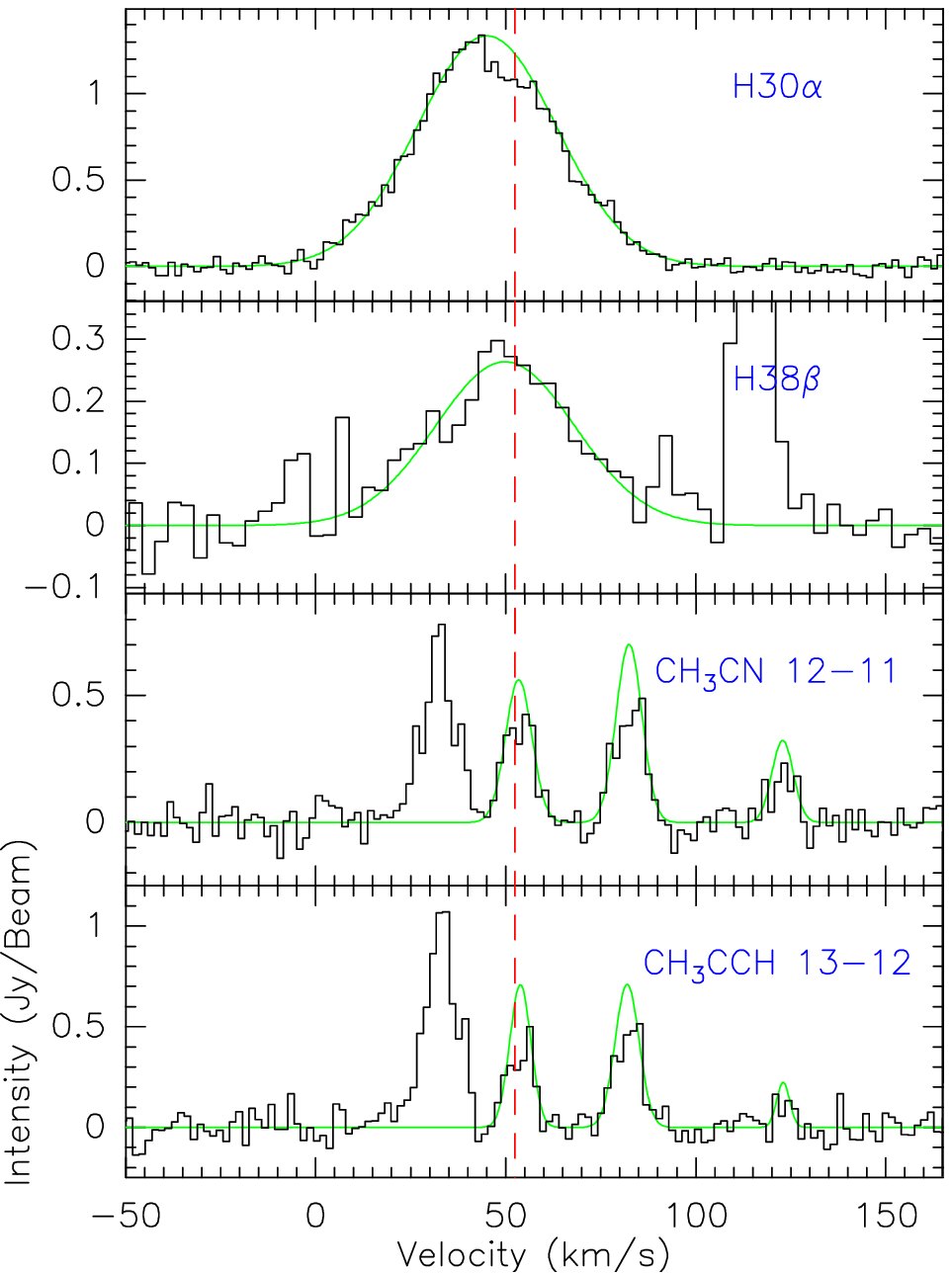}
\end{center}
\caption{ Molecular spectra at the position of the 1.3 mm continuum
peak. The channel spacing is 1.70$~\kms$ in the plots. The red
dashed line denotes the systemic velocity ($V_{sys} = 52.5~\kms$),
crossing the spectra H30$\alpha$, H38$\beta$, CH$_{3}$CN 12(2)-11(2), and
CH$_3$CCH 13(2)-12(2). The green curves are Gaussian fitting lines.}
\label{fig_ch3cn}
\end{figure}

\subsection{VLA} \label{sect:vla}

\begin{center}
\begin{deluxetable*}{lcccccc}
\tabletypesize{\scriptsize} \tablecaption{Continuum emission of
different wavelength} \tablewidth{0pt} \tablehead{
\colhead{Continuum} & \colhead{Right Ascension} & \colhead{Declination}  &  \colhead{Beam Size}  &  \colhead{Deconvolved Size}   &  \colhead{Peak} & \colhead{Total Flux} \\
            & h~~m~~s~~(~$''$)       &  $^\circ~~'~~''~~(~'')$ &  $''\times''$; $^\circ$ &   $''\times''$; $^\circ$  &  Jy/beam        &         Jy    \\
(1)  &   (2)   &  (3)  &  (4) &  (5) & (6) & (7)} \startdata
3.6 cm\tablenotemark{a}      & 18 56 22.563(0.05)  &  2 20 27.660(0.06)     &  2.82$\times$2.38;-12.5      &   1.96$\times$1.35;55.1        &  0.163(0.007)  &   0.234(0.004)          \\
3.6 cm\tablenotemark{b}      & 18 56 22.52         &  2 20 27.0             &  1.05$\times$0.79            &   ---                          &  0.074(0.000)  &   0.197(0.001)          \\
2.0 cm\tablenotemark{b}      & 18 56 22.52         &  2 20 27.3             &  0.57$\times$0.45            &   ---                          &  0.082(0.000)  &   0.242(0.001)          \\
1.3 cm\tablenotemark{a}      & 18 56 22.528(0.01)  &  2 20 27.619(0.02)     &  0.36$\times$0.24;12.6       &   0.66$\times$0.19;23.7        &  0.080(0.006)  &   0.255(0.015)          \\
1.3 mm\tablenotemark{a}      & 18 56 22.554(0.06)  &  2 20 27.705(0.05)     &  3.38$\times$3.05;76.1       &   2.24$\times$1.29;73.6        &  0.811(0.030)  &   1.056(0.017)          \\
0.85 mm\tablenotemark{c}     & 18 56 22.530(0.48)  &  2 20 24.447(0.54)     &  14.00$\times$14.00; 0       &   18.64$\times$14.99;-11.7     &  4.801(0.259)  &   11.712(0.201)          \\
0.45 mm\tablenotemark{c}     & 18 56 22.425(0.51)  &  2 20 25.306(0.56)     &  8.00$\times$8.00; 0         &   11.72$\times$10.04;-17.7     &  21.328(2.003) &   60.701(1.550)          \\
\hline
\enddata
\tablenotetext{a}{SMA and VLA data from this work.}
\tablenotetext{b}{VLA data from \citet{kurt1994}.}
\tablenotetext{c}{JCMT data from archive
(http://www.jach.hawaii.edu/JCMT/archive/). } \label{tab_sed}
\end{deluxetable*}
\end{center}

The 1.3 cm data for G35.58-0.03 were retrieved from the NRAO\footnote{The National
Radio Astronomy Observatory is operated by Associated Universities,
Inc., under cooperative agreement with the National Science
Foundation.} VLA archival database. The project code is AG811.
The 1.3 cm spectral
line observations (NH$_3$ $(J, K)=(2, 2)$ and (3, 3)) are observed
in the VLA-B configuration on 2009 March. 
The phase tracking center
was $\alpha$(J2000) =
18$\mathrm{^h}$56$\mathrm{^m}$22$\rlap.{^{\mathrm{s}}}$500 and
$\delta$(J2000) = 02$^\circ$20$'$27$\rlap.{''}$00. The NH$_3$ $(J,
K)=(2, 2)$ and (3, 3) inversion transitions were observed
simultaneously, using the 2-IF spectral line mode of the correlator,
with 6.25 MHz bandwidth and 127 channels of 49 kHz (0.617$~\kms$)
each. The approximate synthesized beam size (full width at half-power) is 
about $0\dotsec36 \times 0\dotsec24$ with P.A. = 12$\dotdeg$6. The VLA primary 
beam at 23 GHz is $\sim$ 120$''$. The corresponding degrees per flux unit factor 
is DPFU$_{\rm 1.3\,cm} \sim2.5 \times 10^4$ K per Jy beam$^{-1}$. For calibrations, 
J1331+305 was used for flux calibrators, J2253+161 and J1331+305 were used for the
bandpass, and J18517+00355 was observed for the antenna gain and phase
corrections. The flux-density scale was bootstrapped from J1331+305 model 
assuming a flux-density of 2.4059 Jy for 23.72 GHz and 2.3949 Jy for 23.87 GHz. 
And the phase calibrator J18517+00355 has a flux-density of 1.0335$\pm$0.0003 
for 23.72 GHz and 0.8688$\pm$0.0004 for 23.87 GHz. 
The rms noise in the final images is in the range of 
[1, 6]$~\mjyb$ for the line data and $\sim$0.7$~\mjyb$ for the continuum data.

\begin{figure}
\figurenum{4}
\begin{center}
\includegraphics[angle=0,scale=0.85]{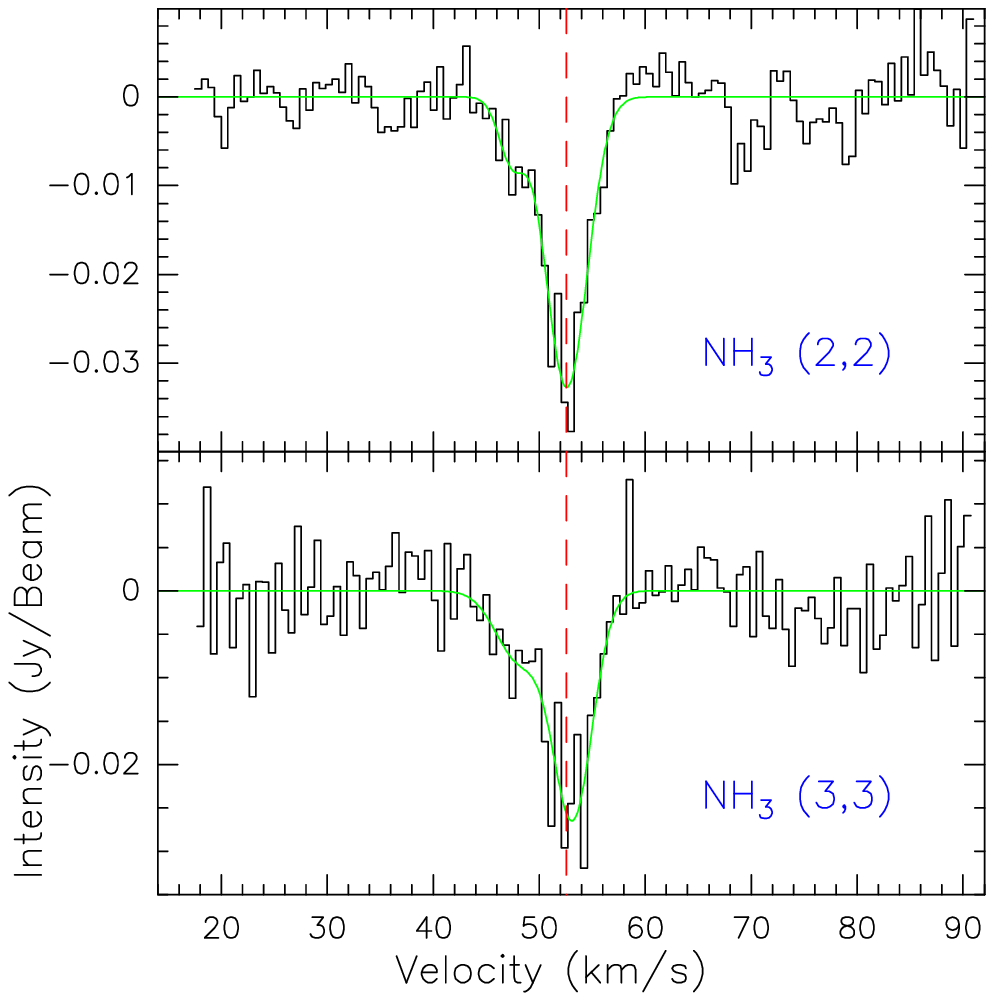}
\end{center}
\caption{  Molecular spectra of NH$_3$ (2, 2) and (3, 3) at the
position of the 1.3 cm continuum peak. The channel spacing is
0.617$~\kms$ in the plots. The red dashed line, crossing the
spectra, denotes the systemic velocity ($V_{sys} = 52.5~\kms$). The
green curves are Gaussian fitting lines.} \label{fig_nh3_spec}
\end{figure}

In addition, the 3.6 cm continuum data were taken from the VLA archive and 
observed on 1998 December with the VLA-C configuration (project code AK477). 
The phase tracking center was $\alpha$(J2000) =
18$\mathrm{^h}$56$\mathrm{^m}$23$\rlap.{^{\mathrm{s}}}$473 and
$\delta$(J2000) = 02$^\circ$20$'$37$\dotsec$76. A bandwidth 50 MHz
was used and centered at 8.4351 (IF1) and 8.4851 (IF2) GHz each.
QSO B1328+307 was used as flux calibrator and QSO B1829-106 was used for gain correction. The rms noise in the final images is  $\sim$6.7$~\mjyb$ for the 3.6 cm continuum data.

All VLA data sets were calibrated using standard procedures in the
AIPS software\footnote{http://www.aips.nrao.edu/index.shtml}. The
1.3 cm continuum was constructed in the $(u, v)$ domain from
line-free channels and was then self-calibrated. The gain solutions
from self-calibration were applied to the line data. The calibrated
data were exported to GILDAS and MIRIAD for further processing and
imaging.

\section{Observational Results} \label{sect:result}

\subsection{Spectra}

Figure \ref{fig_usb_lsb} shows the SMA 4 GHz continuum-free spectra
extracted from the line-cubes at the position of the 1.3
mm continuum peak smoothed to 4.25$~\kms$ channel
spacing. The original spectral resolution 
of $0.85~\kms$ was used for the data analysis. For the spectral 
identification, we checked the line
observations of G20.08-0.14N \citep{galv2009}. Then, the rest frequencies were mainly assigned with
spectral catalogs of JPL \citep{pick1998}, CDMS \citep{mull2005},
and SPLATALOGUE\footnote{http://www.splatalogue.net/}.  Finally, we
identified 25 transitions from 8 molecular species and their
isotopologues. The relevant line parameters are listed in Table
\ref{tab_lines}.

\begin{figure*}
\figurenum{5}
\begin{center}
\includegraphics[angle=-90,scale=0.35]{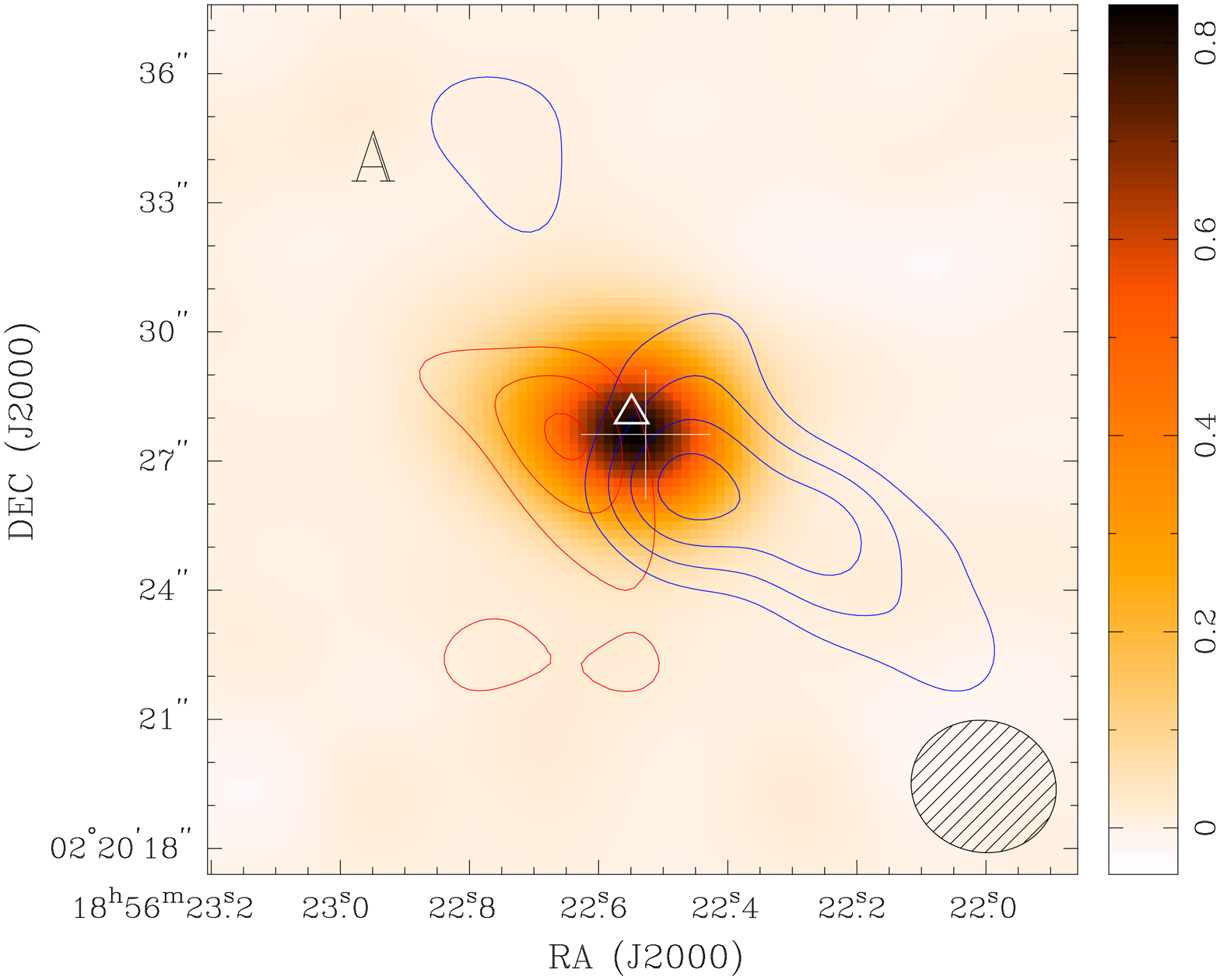}
\includegraphics[angle=-90,scale=0.35]{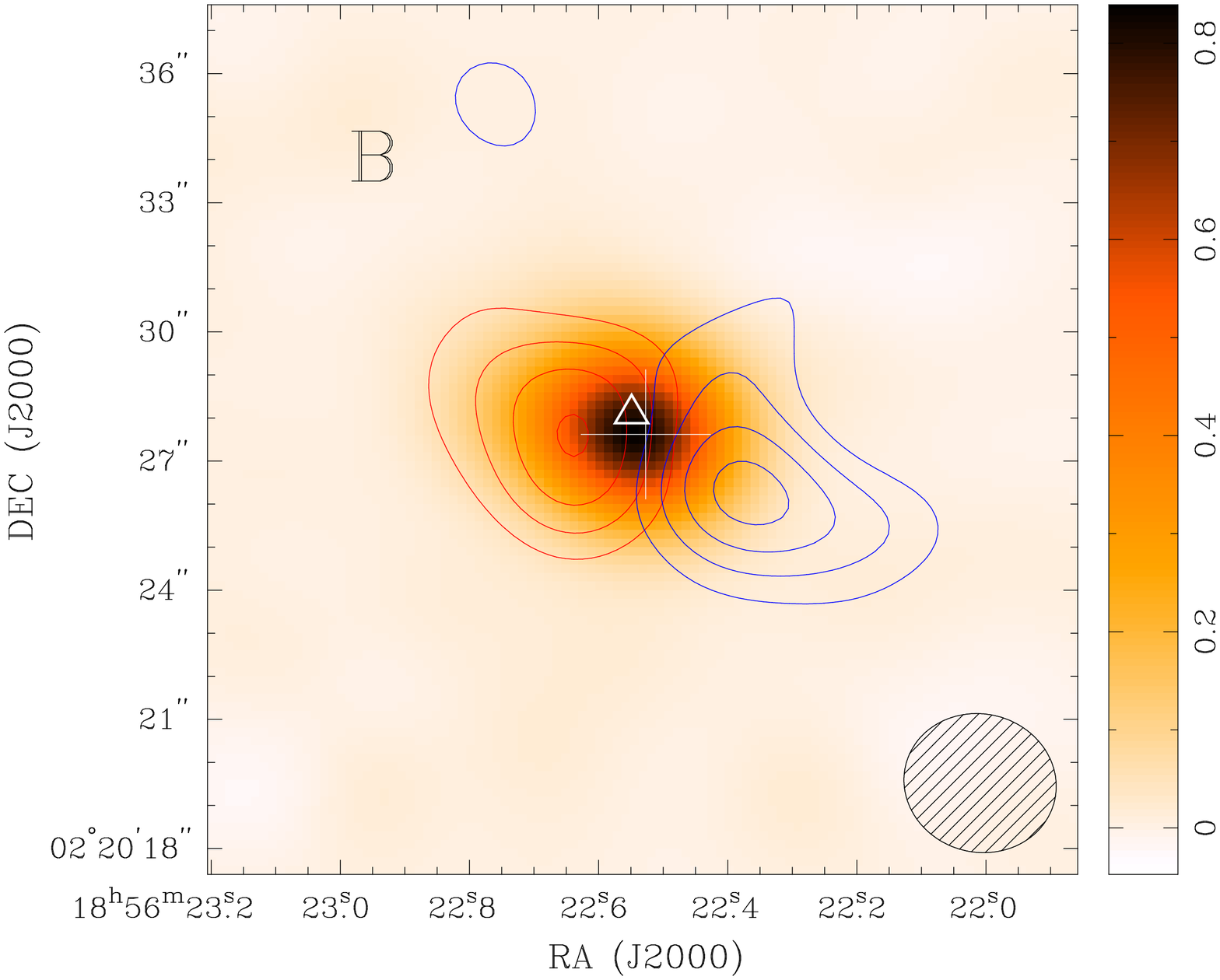}
\includegraphics[angle=-90,scale=0.35]{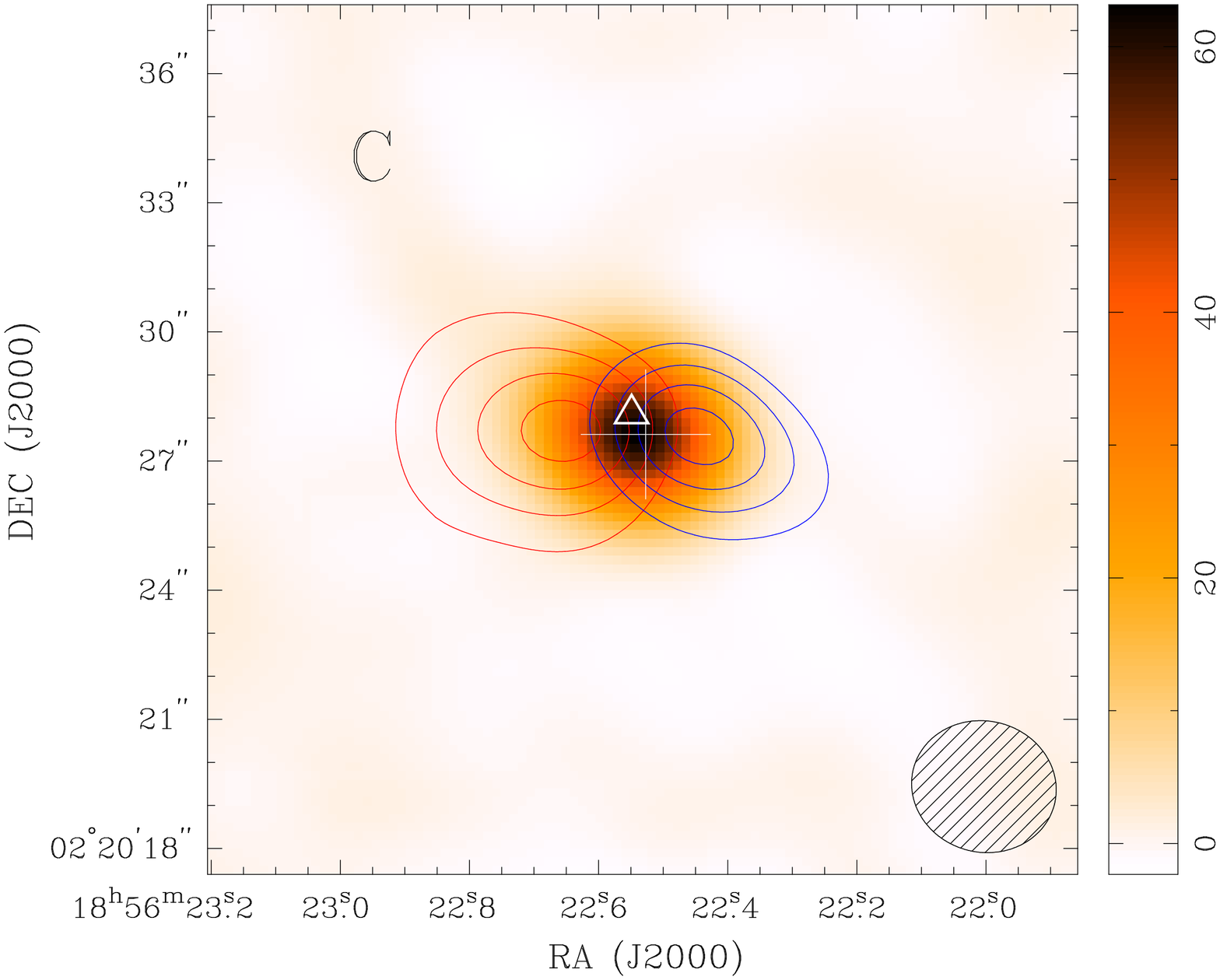}
\includegraphics[angle=-90,scale=0.35]{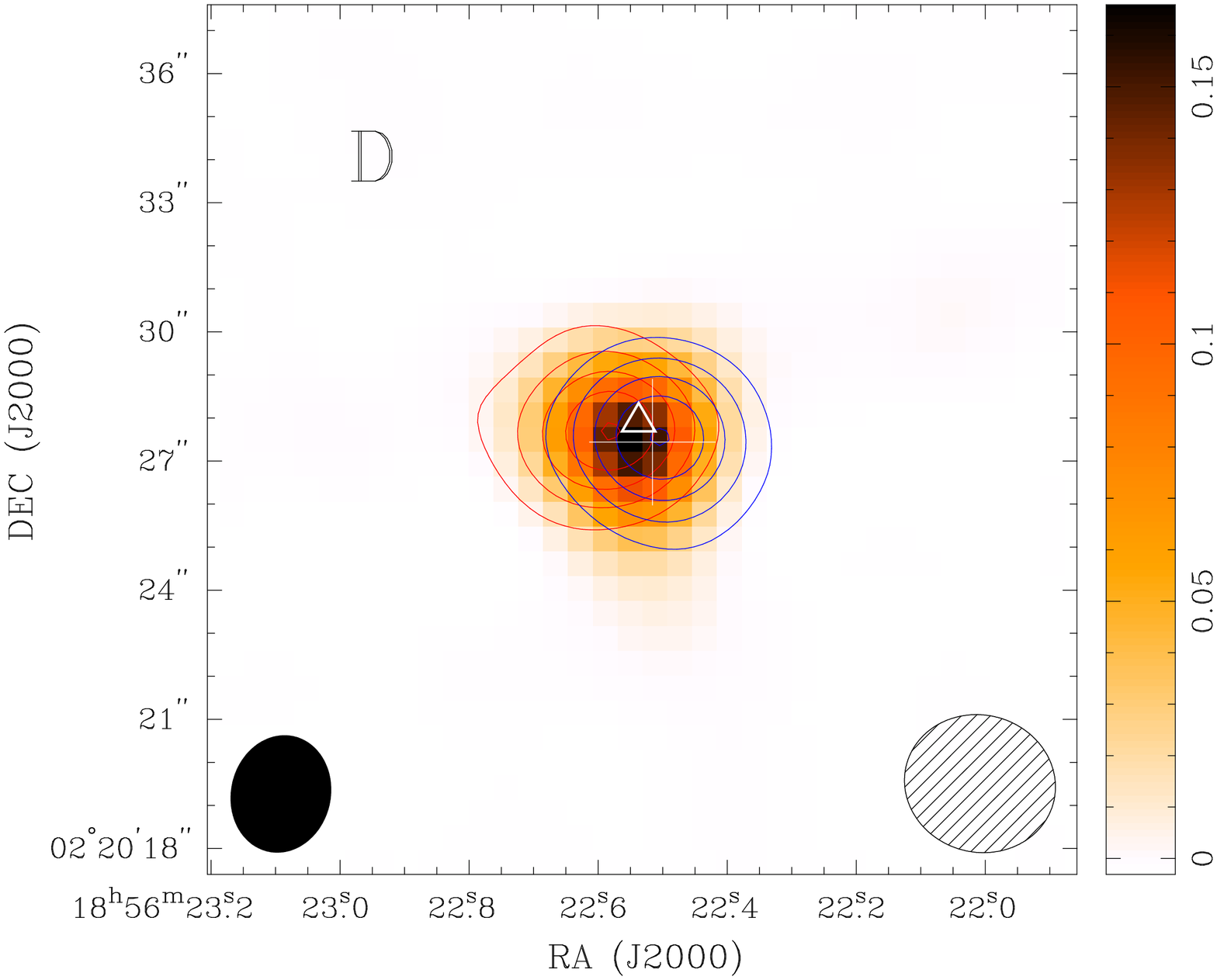}
\end{center}
\caption{ (A) Velocity-integrated contours of CO (2-1) superimposed
on the 1.3 mm continuum image. The blue contours of [45, 50]$~\kms$
are 3, 4, 5, 6 $\times 1.70~\jybkms$, and the red contours of [59,
65]$~\kms$ are 3, 4, 5 $\times 0.79~\jybkms$, respectively. The 1.3 mm
continuum peak is 0.811$~\jyb$. (B) Velocity-integrated contours of
$^{13}$CO (2-1) superimposed on the 1.3 mm continuum image. The blue
contours of [40, 50]$~\kms$ are 3, 4, 5, 6 $\times 0.18~\jybkms$, and
the red contours of [56, 66]$~\kms$ are 3, 4, 5, 6 $\times
1.28~\jybkms$, respectively. The 1.3 mm continuum peak is 0.811$~\jyb$.
(C) Velocity-integrated contours of $^{13}$CS (5-4) superimposed on
the H30$\alpha$ velocity-integrated image. The blue contours of [46, 51]$~\kms$
are 3, 5, 7, 9 $\times 0.41~\jybkms$, and the red contours of [54,
60]$~\kms$ are 3, 5, 7, 9 $\times 0.57~\jybkms$, respectively. The
H30$\alpha$ flux peak is 61.37$~\jybkms$ for integration range [0, 90]$~\kms$. (D) Velocity-integrated
contours of SO$_2$ $11(1,11)-10(0,10)$ superimposed on the 3.6 cm
continuum image. The blue contours of [46, 51]$~\kms$ are 3, 5, 7,
9, 11 $\times 0.62~\jybkms$, and the red contours of [54, 60]$~\kms$
are 3, 5, 7, 9, 11 $\times 0.29~\jybkms$, respectively. The 3.6 cm
continuum peak is 0.163$~\jyb$. The white cross indicates the position
of 1.3 cm continuum peak. The white triangle indicates the position
of water and OH masers. The synthesized beam of 1.3 mm and 1.3 cm
data are in hatched and solid ellipses, respectively.}
\label{fig_br}
\end{figure*}

Figures \ref{fig_spectra} and \ref{fig_ch3cn} show Gaussian
fits of several lines including CO (2-1), $^{13}$CO (2-1), $^{13}$CS
(5-4), OCS (19-18), SO$_2$ 11(1,11)-10(0,10), $^{34}$SO$_2$
22(2,20)-22(1,21), $^{34}$SO$_2$ 13(2,12)-13(1,13), H30$\alpha$, H38$\beta$,
CH$_3$CN (12-11), and CH$_3$CCH (13-12). The systemic velocity
$V_{sys}\approx52.5~\kms$ was derived from the mean value of
the velocities from Gaussian fits. The spectra were extracted from the
position of the 1.3 mm continuum peak. The channel spacings are 0.85
and 1.70 $\kms$ for Figures \ref{fig_spectra} and \ref{fig_ch3cn},
respectively. Since the spectral profiles are absorbed strongly at
the line center, we assumed that the spectral features are from
self-absorption and missing flux, and then masked the absorption dip to make Gaussian
fitting using CLASS software. The CH$_3$CN 12(0)-11(0) with
12(1)-11(1) and CH$_3$CCH 13(0)-12(0) with 13(1)-12(1) are blended, so they are not fitted. The fitted line parameters are listed in Table \ref{tab_lines}.

Figure \ref{fig_nh3_spec} shows the VLA absorption spectra of NH$_3$
(2, 2) and (3, 3) at the position of the 1.3 cm continuum peak. The
channel spacing is 0.617 $\kms$. We only fitted the main lines of
NH$_3$ with Gaussian profiles, because the satellite lines are too
noisy to be identified. The fitted line parameters are listed in
Table \ref{tab_lines}.

In Table \ref{tab_lines}, Column 1 lists the names of
molecular species. Columns 2 -- 3 list the transitions and
rest frequencies of the molecules, respectively. Columns 4 -- 7 list the
Gaussian fitting results including integration flux, central line
velocity, full width at half-maximum (FWHM), and peak intensity,
respectively. Uncertainties are shown in parentheses following each
Gaussian fitting result.

\subsection{Moment Maps}

\begin{figure*}
\figurenum{6}
\begin{center}
\includegraphics[angle=-90,scale=0.36]{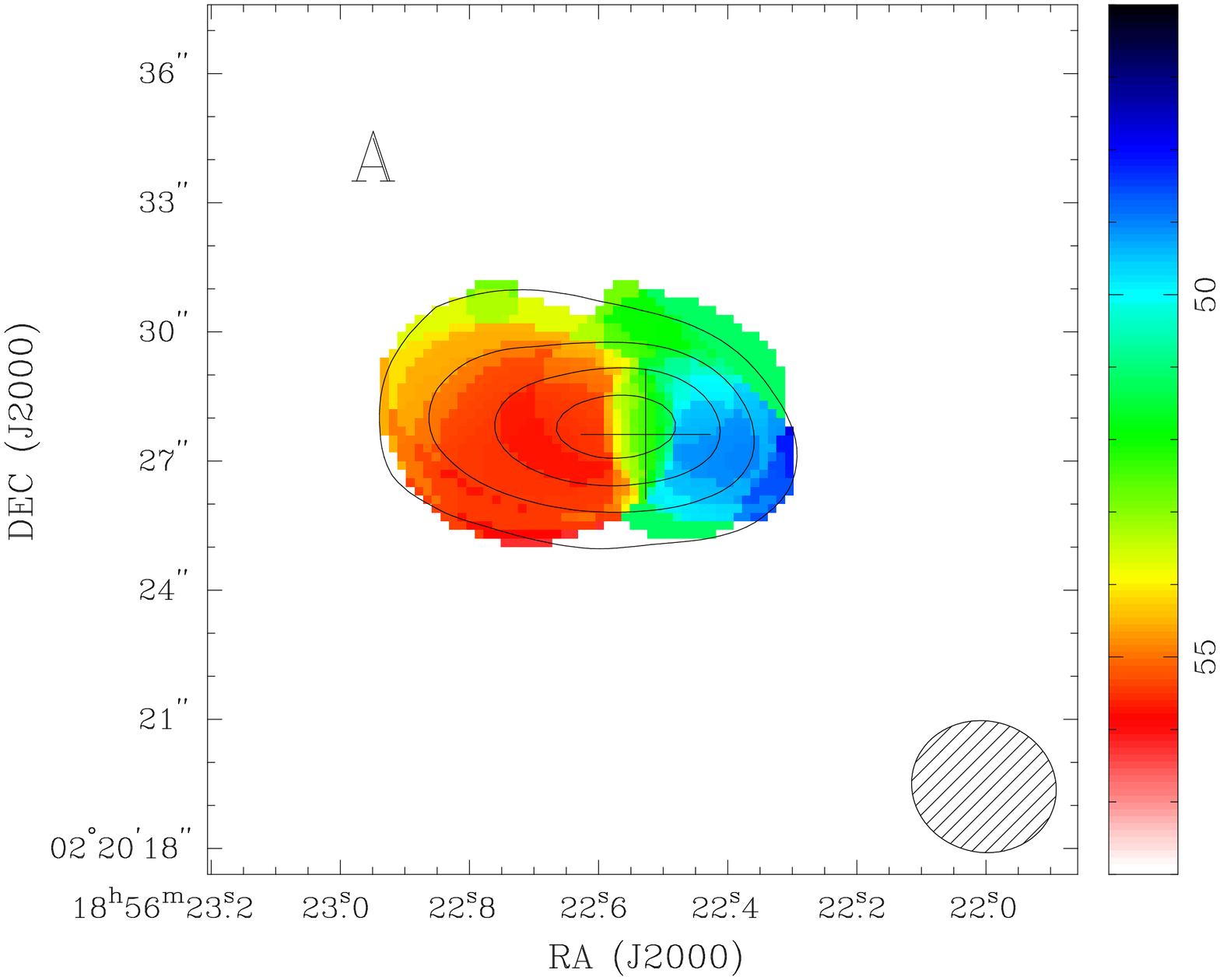}
\includegraphics[angle=-90,scale=0.36]{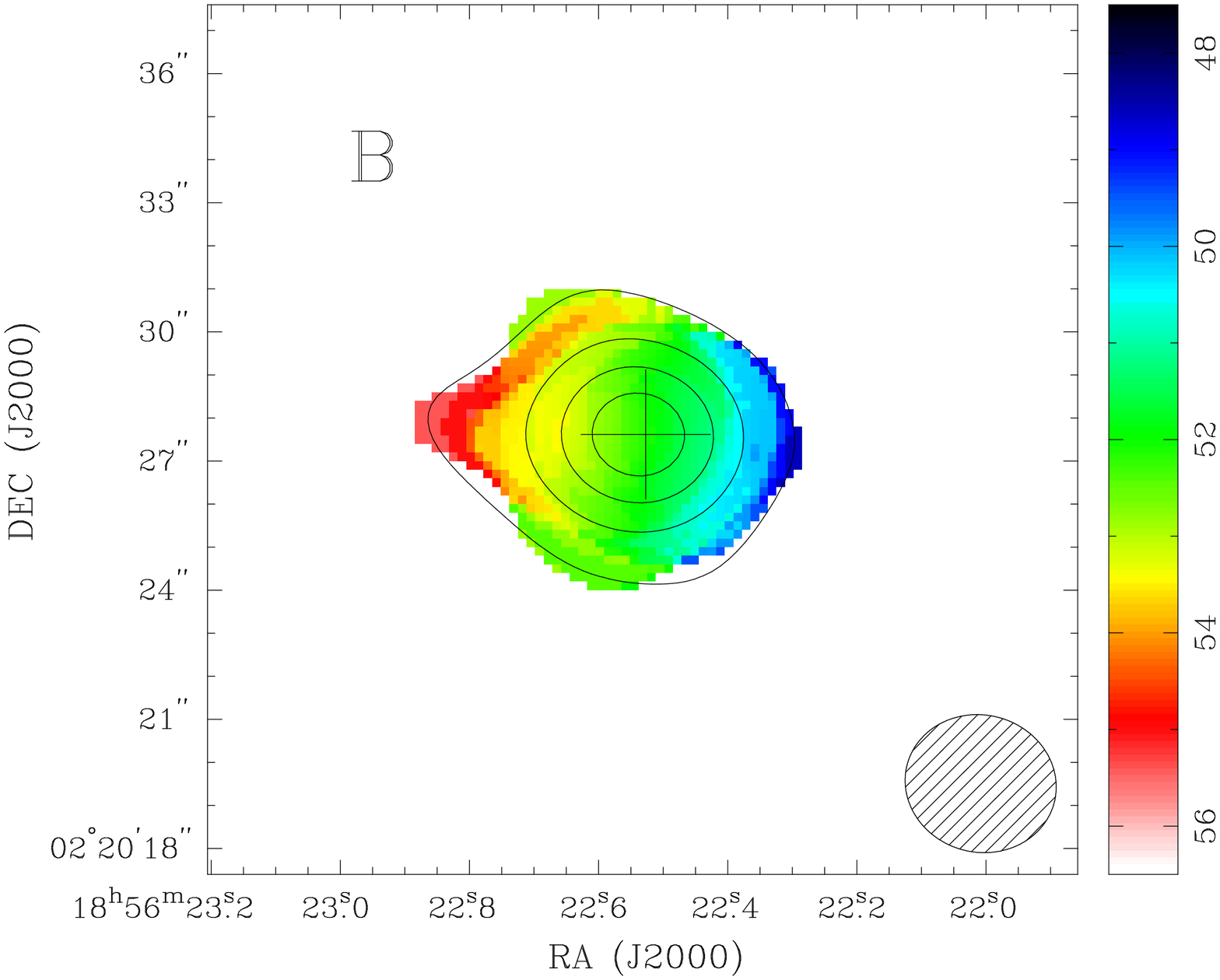}
\includegraphics[angle=-90,scale=0.36]{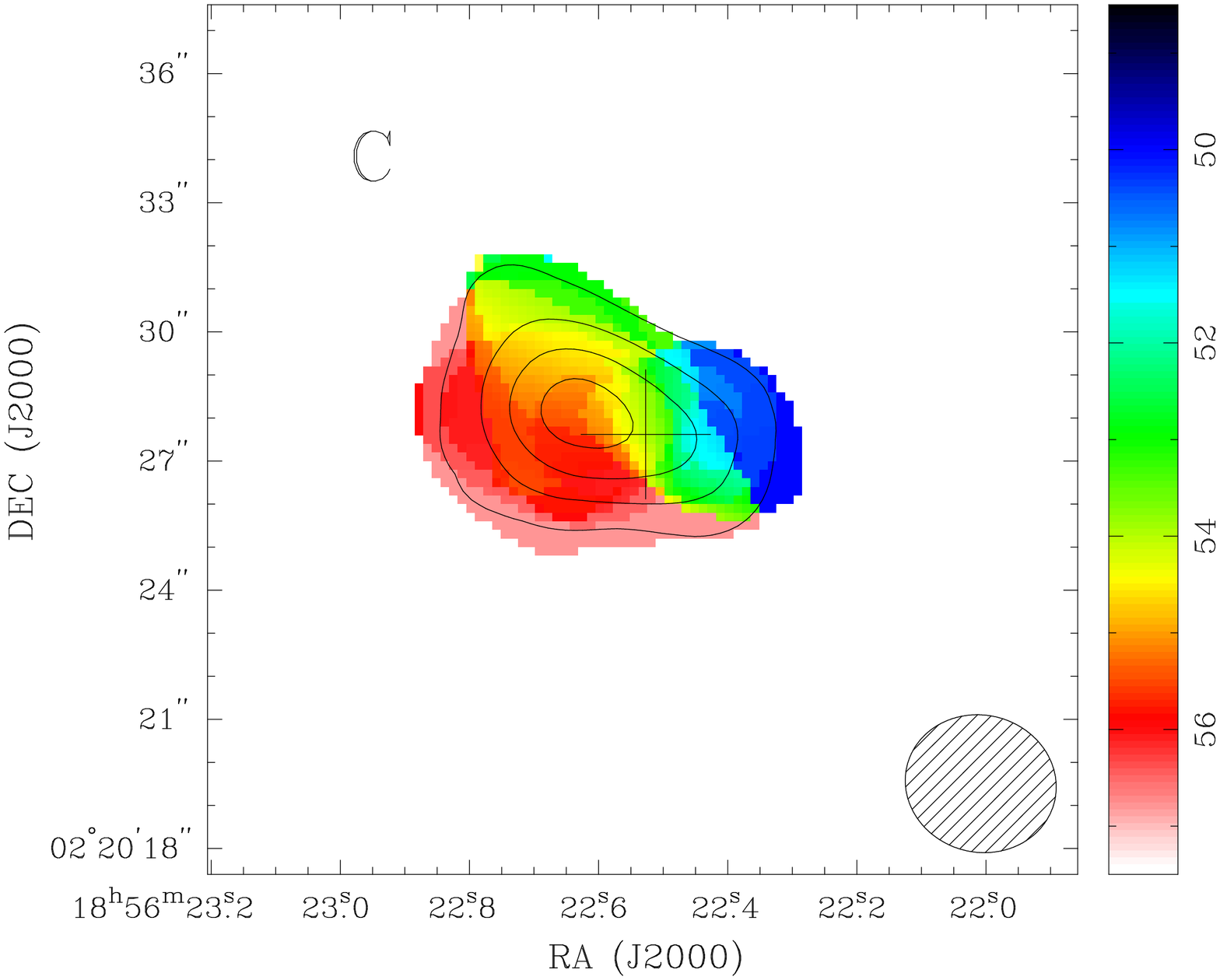}
\includegraphics[angle=-90,scale=0.36]{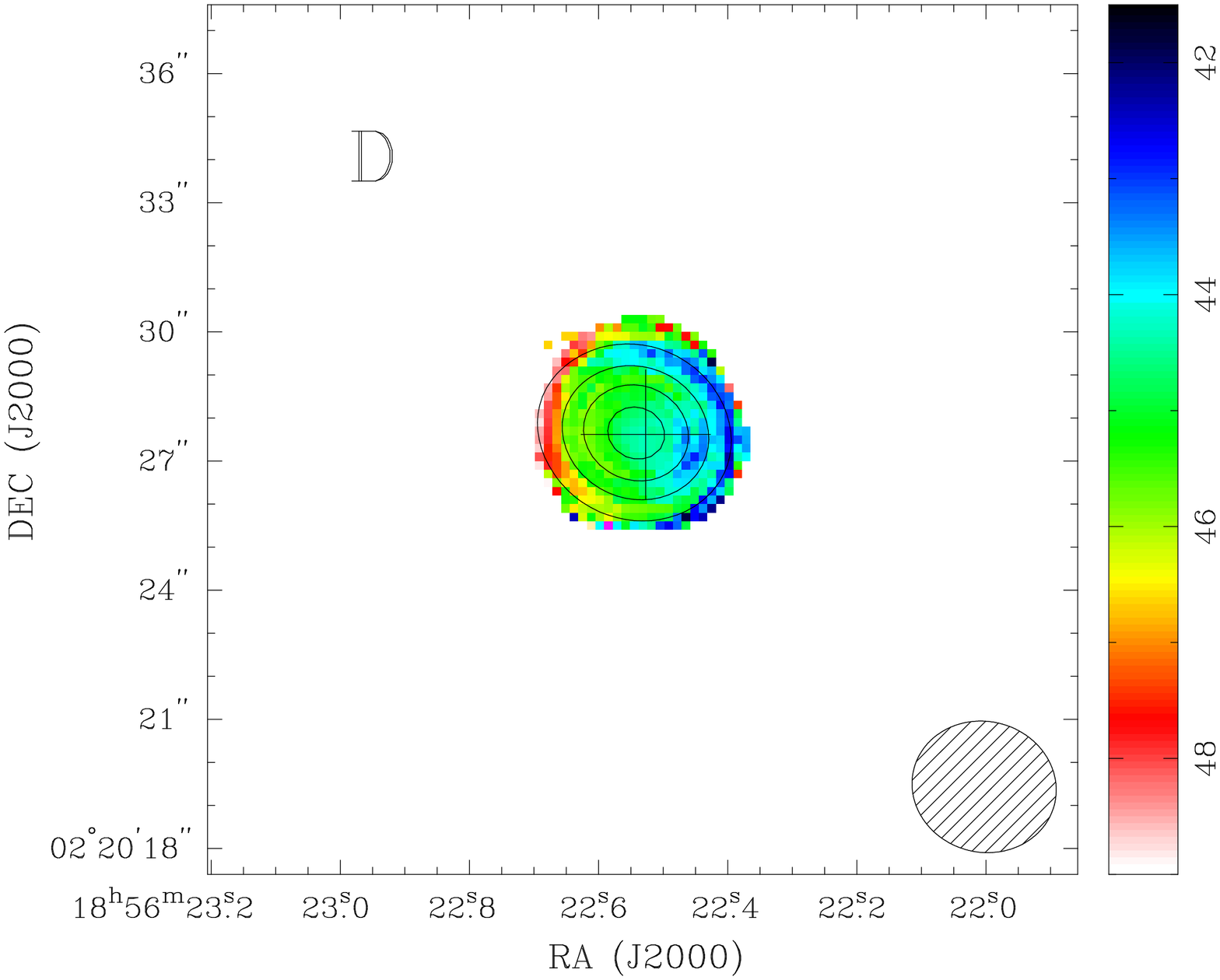}
\end{center}
\caption{ Velocity-integrated emission (moment 0, {\it contours})
superimposed on the intensity-weighted mean velocity (moment 1, {\it
color scale}) maps. (A) Contours of $^{13}$CS (5-4) are 3, 5, 7, 9
$\times 0.87~\jybkms$; (B) Contours of SO$_2$ 11(1,11)-10(0,10) are
3, 11, 19, 27 $\times 0.46~\jybkms$; (C) Contours of CH$_3$CCH
13(2)-12(2) are 3, 5, 7, 9 $\times 0.41~\jybkms$; (D) Contours of
H30$\alpha$ are 3, 5, 7, 9 $\times 6.32~\jybkms$. The black cross
indicates the position of 1.3 cm continuum peak. The units of color
bar are in$~\kms$. } \label{fig_sma_mom1}
\end{figure*}

Figure \ref{fig_br} shows (A) velocity-integrated contours of CO
(2-1) superimposed on the 1.3 mm continuum image, (B)
velocity-integrated contours of $^{13}$CO (2-1) superimposed on the
1.3 mm continuum image, (C) velocity-integrated contours of
$^{13}$CS (5-4) superimposed on the H30$\alpha$ emission image, and
(D) velocity-integrated contours of SO$_2$ 11(1,11)-10(0,10)
superimposed on the 3.6 cm continuum image. The blue and red
contours show integrated emission from blue and red wings, and the
velocity ranges are presented in the caption. The peak intensities
of 1.3 mm continuum, H30$\alpha$ emission, and 3.6 cm continuum are
0.811 $\jyb$, 61.37 $\jybkms$, and 0.163 $\jyb$, respectively. The white cross indicates
the position of the 1.3 cm continuum peak. The relevant coordinates and
beam sizes are listed in Table \ref{tab_sed}. The white triangle
indicates the positions of a water ($\alpha$(J2000) =
18$\mathrm{^h}$56$\mathrm{^m}$22$\rlap.{^{\mathrm{s}}}$550 and
$\delta$(J2000) = 02$^\circ$20$'$28$\dotsec$100) and OH maser
($\alpha$(J2000) =
18$\mathrm{^h}$56$\mathrm{^m}$22$\rlap.{^{\mathrm{s}}}$540 and
$\delta$(J2000) = 02$^\circ$20$'$28$\dotsec$100)
\citep{fors1989,debu2005}, but no methanol masers have been detected \citep{casw1995}.

\begin{figure*}
\figurenum{7}
\begin{center}
\includegraphics[angle=-90,scale=0.36]{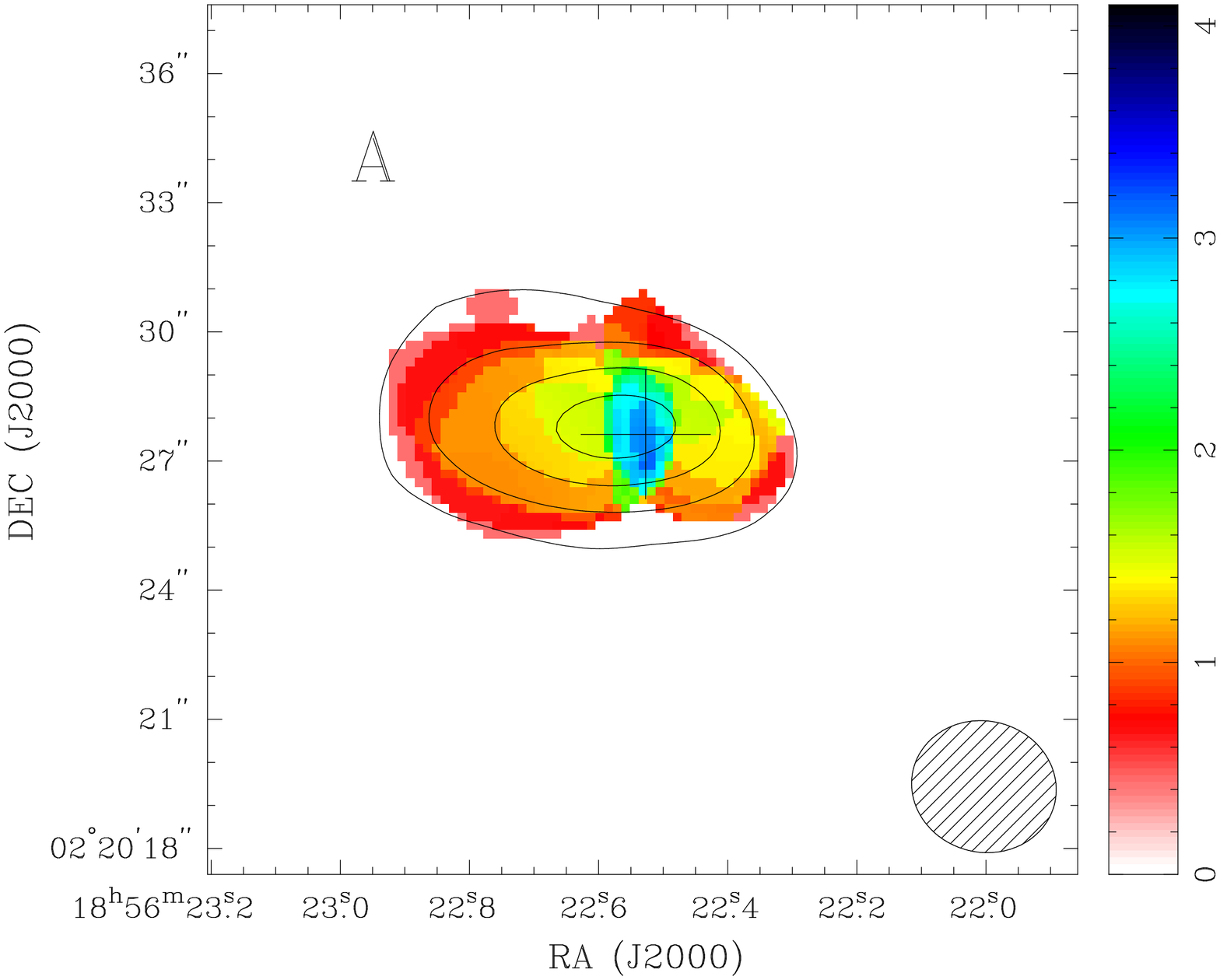}
\includegraphics[angle=-90,scale=0.36]{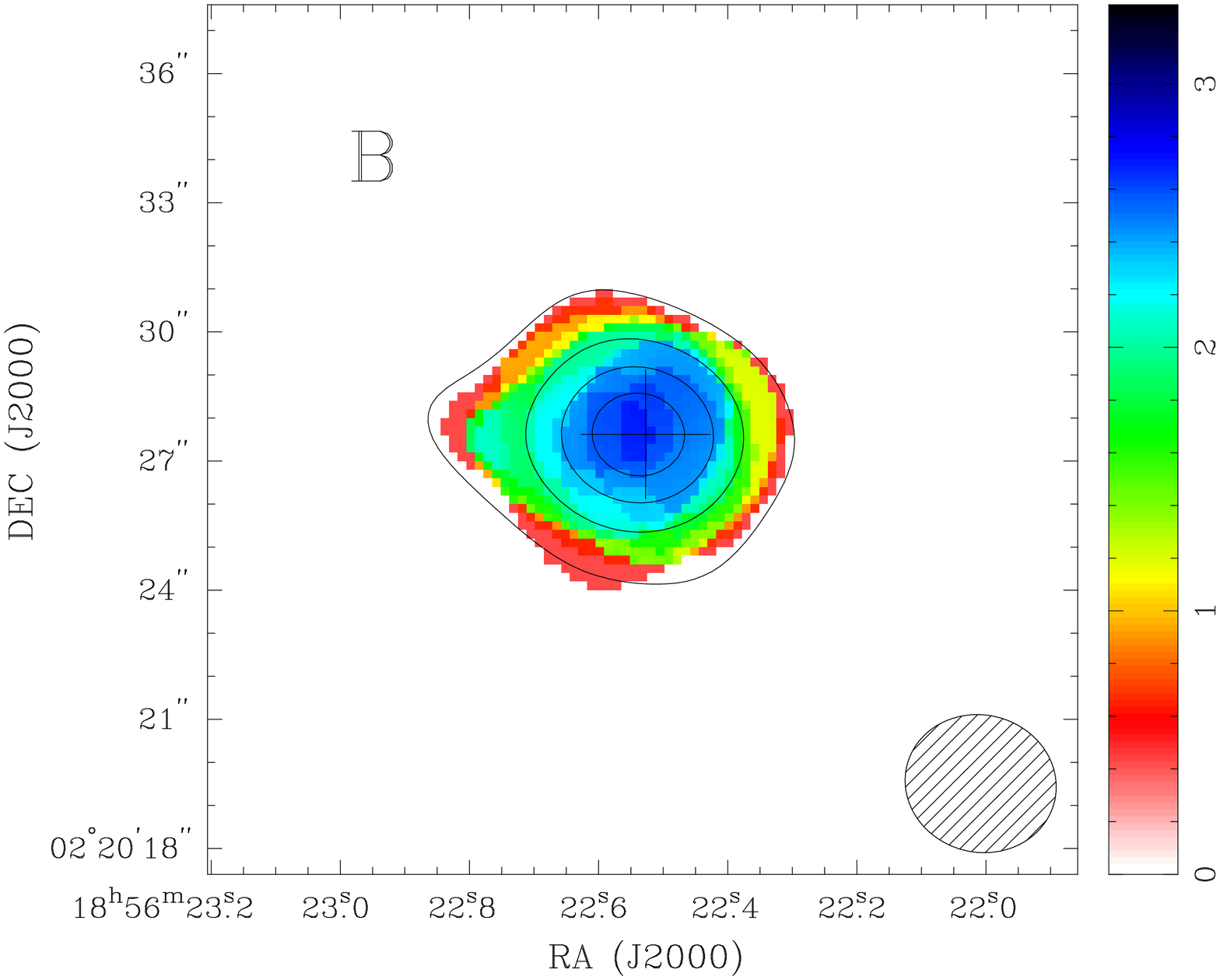}
\includegraphics[angle=-90,scale=0.36]{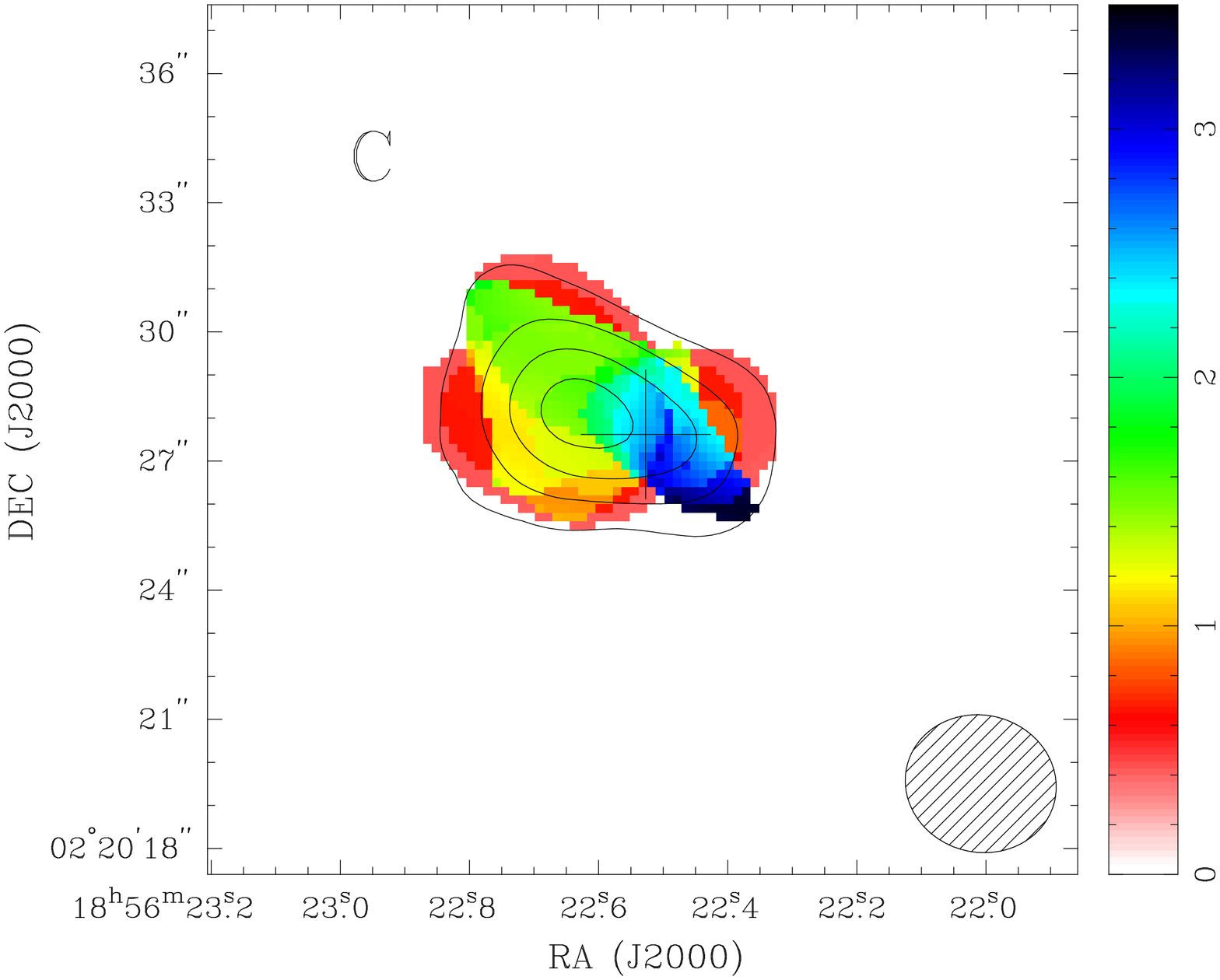}
\includegraphics[angle=-90,scale=0.36]{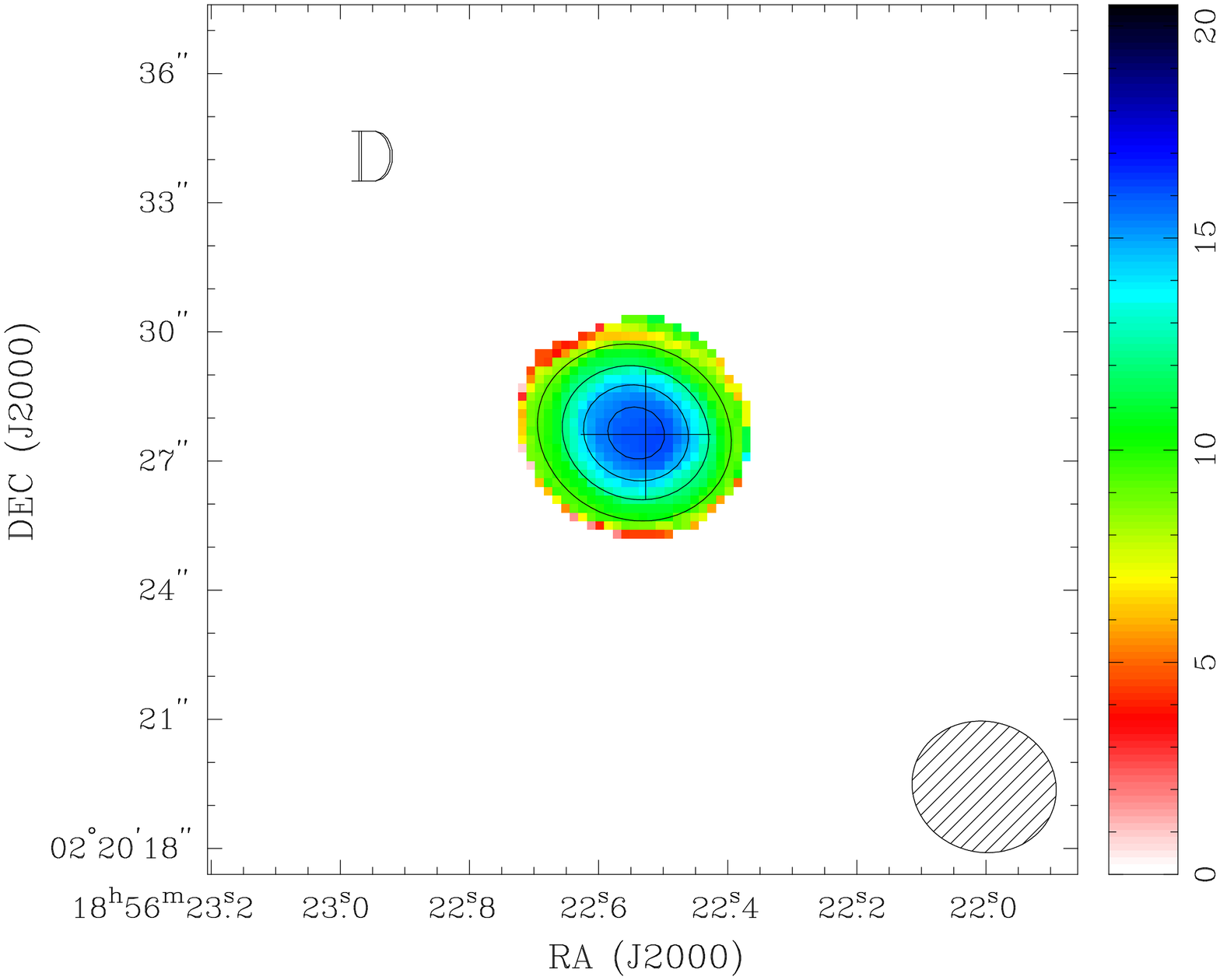}
\end{center}
\caption{ Velocity-integrated emission (moment 0, {\it contours})
superimposed on the velocity dispersion (moment 2, {\it color
scale}) with respect to moment-1 velocity maps. (A) Contours of
$^{13}$CS (5-4) are 3, 5, 7, 9 $\times 0.87~\jybkms$; (B) Contours
of SO$_2$ 11(1,11)-10(0,10) are 3, 11, 19, 27 $\times 0.46~\jybkms$;
(C) Contours of CH$_3$CCH 13(2)-12(2) are 3, 5, 7, 9 $\times 0.41
~\jybkms$; (D) Contours of H30$\alpha$ are 3, 5, 7, 9 $\times 6.32
~\jybkms$. The black cross indicates the position of 1.3 cm
continuum peak. The units of color bar are in$~\kms$. }
\label{fig_sma_mom2}
\end{figure*}

Figure \ref{fig_sma_mom1} shows the velocity-integrated emission
(moment 0, contours) superimposed on the intensity-weighted mean
velocity (moment 1, color scale) maps. Figure \ref{fig_sma_mom2}
shows the velocity-integrated emission (moment 0, contours)
superimposed on the velocity dispersion (moment 2, color scale) with
respect to moment-1 velocity maps. The mapped molecular species
include $^{13}$CS (5-4), SO$_2$ 11(1,11)-10(0,10), CH$_3$CCH
13(2)-12(2), and H30$\alpha$.

Figure \ref{fig_nh3_mom0} shows the velocity-integrated contours of
NH$_{3}$ (2, 2) and (3, 3) superimposed on the 1.3 cm continuum
image. The velocity ranges are presented in the figure caption. The
peak intensity of the 1.3 cm continuum is 80.29 $\mjyb$. The white
cross indicates the position of 1.3 cm continuum peak. The relevant
coordinates and beam sizes are listed in Table \ref{tab_sed}. The
hollow and filled-black triangles indicate the positions of water
maser ($\alpha$(J2000) =
18$\mathrm{^h}$56$\mathrm{^m}$22$\rlap.{^{\mathrm{s}}}$550 and
$\delta$(J2000) = 02$^\circ$20$'$28$\dotsec$100) and OH maser
($\alpha$(J2000) =
18$\mathrm{^h}$56$\mathrm{^m}$22$\rlap.{^{\mathrm{s}}}$540 and
$\delta$(J2000) = 02$^\circ$20$'$28$\dotsec$100)
\citep{fors1989,debu2005}. The black ``$\times$" symbols indicate
two separated \HII regions: G35.578-0.030 to the west, and
G35.578-0.031 to the east \citep{kurt1994,kurt1999,debu2005},
however, our work shows that from the 1.3 cm continuum the eastern 
G35.578-0.031 is very faint. In fact, the G35.578-0.030 is the main research
object in this work.

Figure \ref{fig_nh3_mom1} shows the velocity-integrated contours of
NH$_3$ (2, 2) and (3, 3) superimposed on the intensity-weighted mean
velocity (moment 1, color scale) maps. Figure \ref{fig_nh3_mom2}
shows the velocity-integrated contours of NH$_{3}$ (2, 2) and (3, 3)
superimposed on the velocity dispersion (moment 2, color scale) with
respect to moment-1 velocity maps.

\subsection{PV Diagrams}

Figure \ref{fig_pv} shows the PV diagrams of 1.3
mm observations with P.A. = 90$^\circ$ at the
position of the SMA 1.3 mm continuum peak. The mapped molecular
species include CO (2-1), $^{13}$CO (2-1), $^{13}$CS (5-4), SO$_2$
11(1,11)-10(0,10), H30$\alpha$, CH$_3$CN 12(2)-11(2), CH$_3$CN
12(3)-11(3), CH$_3$CCH 13(2)-12(2), and CH$_3$CCH 13(2)-12(2). The
unit of contours is in Kelvin (DPFU$_{\rm 1.3\,mm} \sim 2.22$ 
K per Jy beam$^{-1}$). The PV diagrams are useful to understand the infall, 
outflow, and/or rotation motions of molecular gas.

\section{Results and Discussions} \label{sect:disc}

\subsection{Spectral Energy Distribution} \label{sect:sed}

\begin{figure}
\figurenum{12}
\begin{center}
\includegraphics[angle=0,scale=0.50]{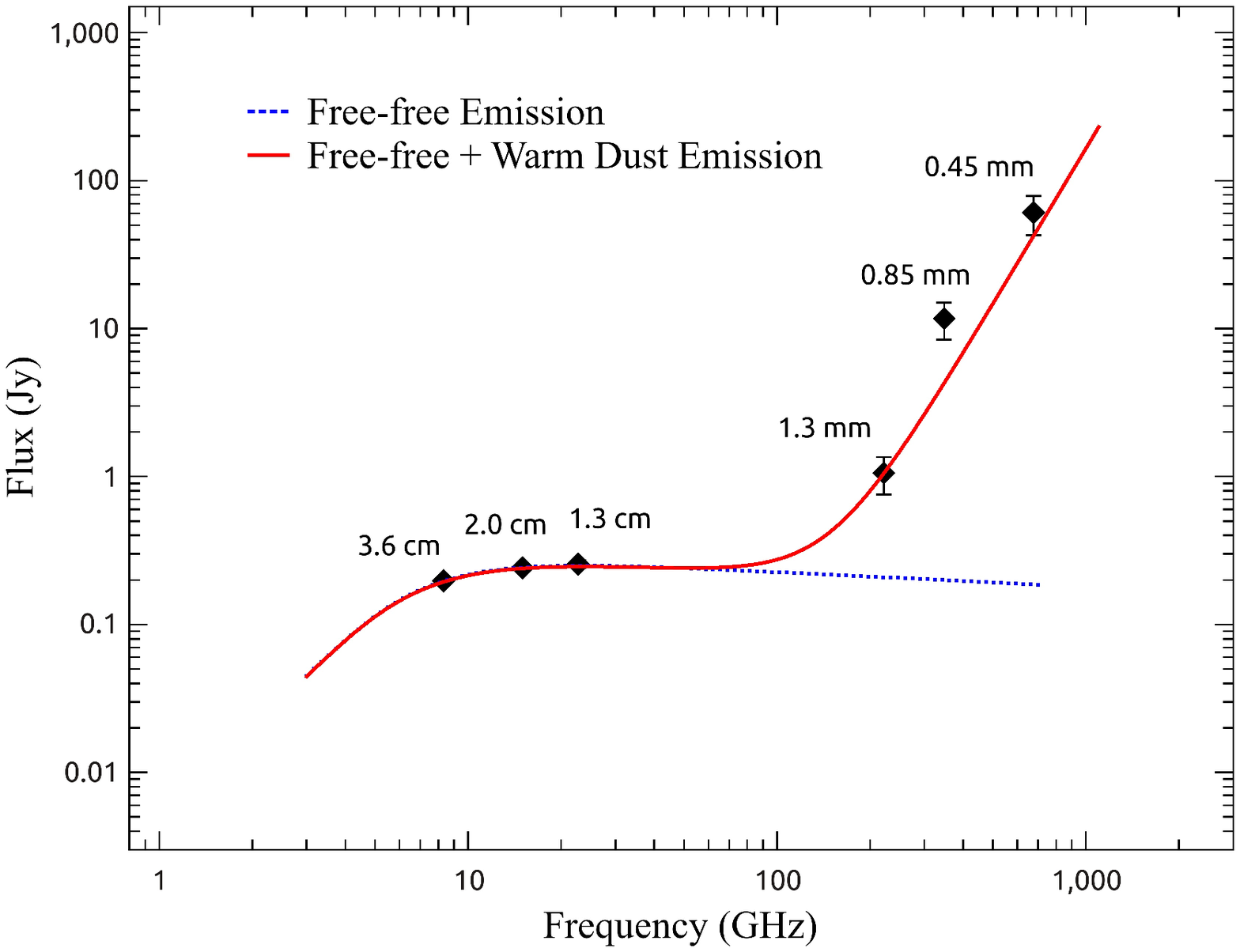}
\caption{ SED for HC \HII G35.58-0.03 combining 3.6 cm, 2.0 cm, 1.3
cm, 1.3 mm, 0.85 mm, and 0.45 mm continuum data. The continuum total
fluxes are listed in Table \ref{tab_sed}. The respective error bars
were indicated in each data point. The dotted blue line is the SED 
fitting from free-free emission component, while the solid red line is 
the sum of the free-free emission and warm dust emission.} \label{fig_sed}
\end{center}
\end{figure}

Figure \ref{fig_sed} exhibits the SED
of the HC \HII region G35.58-0.03 combining 3.6 cm, 2.0 cm, 1.3 cm, 1.3
mm, 0.85 mm, and 0.45 mm continuum data. Their fluxes are listed in
Table \ref{tab_sed}. The modeling equations are described in \citet{shih2010}, 
one of which describes the free-free emission of the ionized gas component, 
another the thermal emission from warm dust. The SED fitting results show 
that about 25\% of the 1.3 mm continuum flux is contributed by the ionized gas component
and about 75\% by the warm dust component. Comparing the dense core seen in the high 
resolution 1.3 cm continuum with the low resolution 1.3 mm continuum data, we suggest 
that there exists a thick dust envelope surrounding the HC \HII core. 
The dust emission arises from a much larger region than the free-free emission.
Generally, the emission at the short wavelength
side traces the dust in a disk/envelope system with a steep spectral
index of $\alpha_{\rm dust}\gtrsim2$, while the emission at the long wavelength
side is from ionized gas of thermal radio jet with a flat spectral
index of $\alpha_{\rm gas}\lesssim1$ \citep{reyn1986,angl1998,choi2012}.
In this work, we obtain the two spectral indexes with $\alpha_{\rm dust}=3.78\pm0.03$ 
and $\alpha_{\rm gas}=0.32\pm0.04$, which indicates a 
density gradient within the ionized gas \citep{fran2000}.  
In addition, considering the 1.3 mm continuum flux contribution from the ionized
gas component to be low, we derive an upper limit of 55\% for the ionized gas 
component using the spectral index $\alpha_{\rm gas}=0.32\pm0.04$.

\begin{figure}
\figurenum{8}
\begin{center}
\includegraphics[angle=-90,scale=0.35]{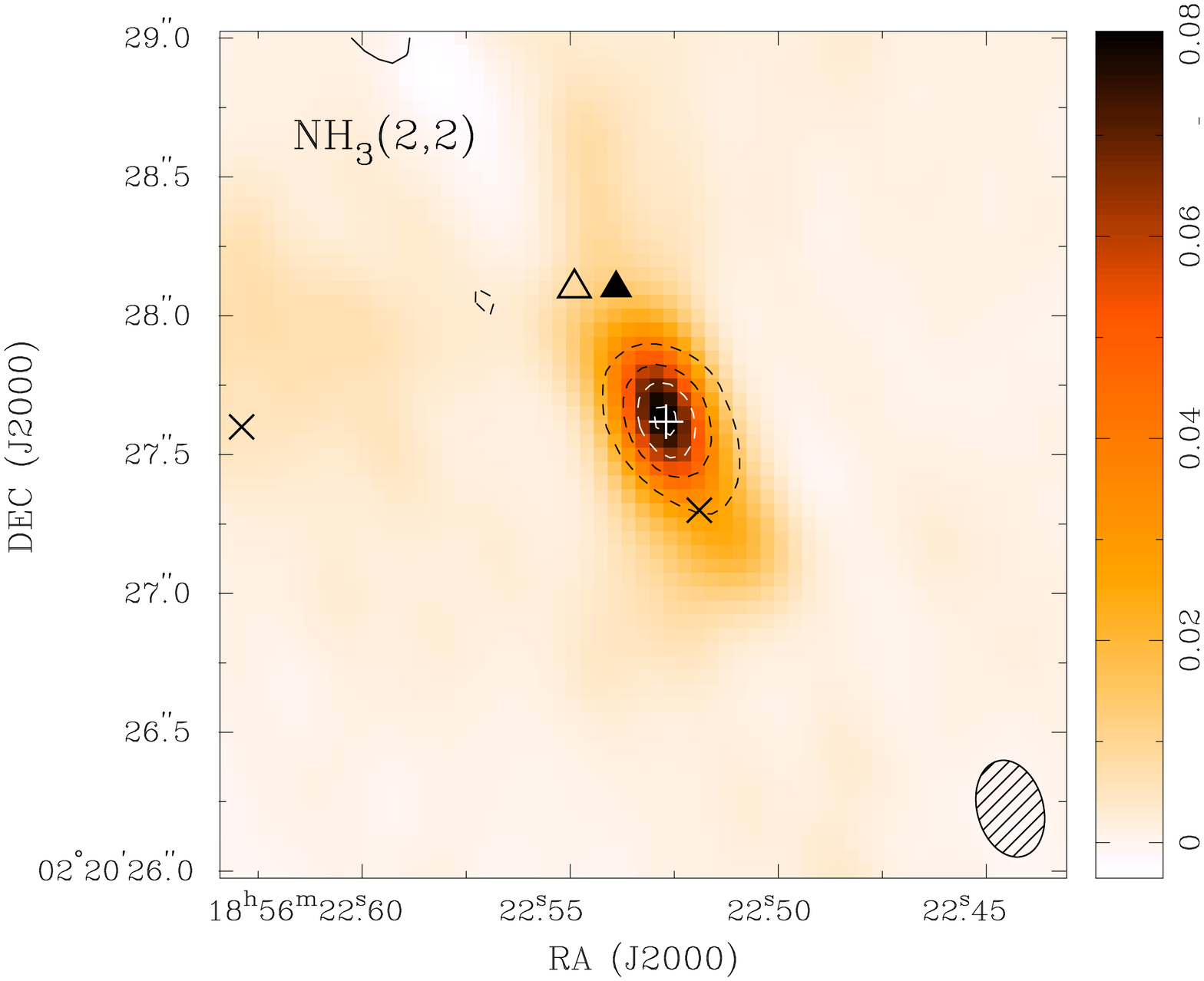}
\includegraphics[angle=-90,scale=0.35]{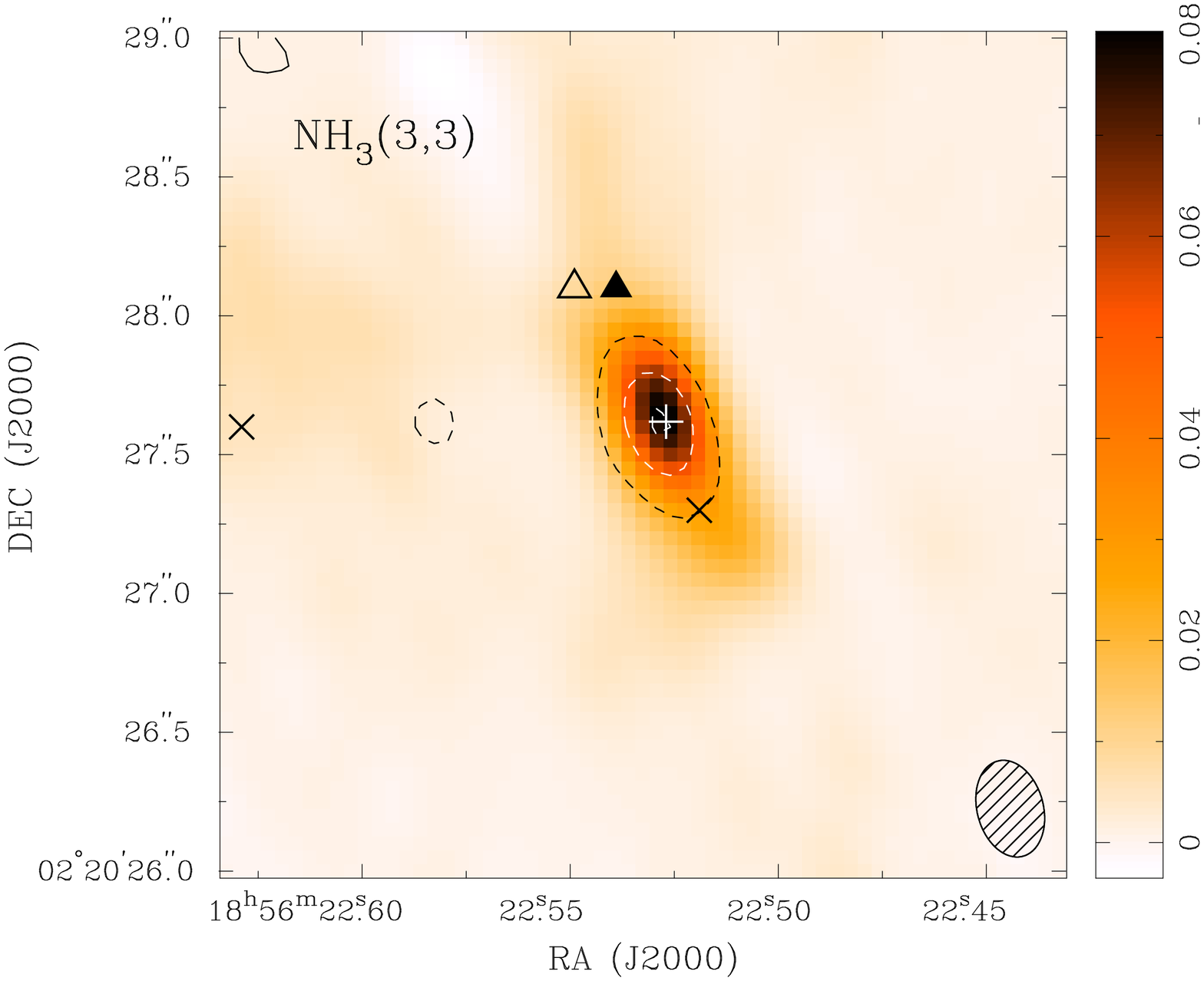}
\caption{ Velocity-integrated contours of NH$_3$ (2, 2) and (3, 3)
superimposed on the 1.3 cm continuum image. Contours of NH$_3$ (2,
2) are -21, -15, -9, -3, 3 $\times 7.64~\mjybkms$; Contours of
NH$_3$ (3, 3) are -15, -9, -3, 3 $\times 10.99~\mjybkms$. The 1.3 cm
continuum peak is 80.29$~\mjyb$. The white cross indicates the
position of 1.3 cm continuum peak. The hollow and fill-black
triangles indicate the positions of water and OH masers. The black
symbols $``\times"$ indicate two UC \HII regions from
\citet{kurt1994,kurt1999}. } \label{fig_nh3_mom0}
\end{center}
\end{figure}

In Figure \ref{fig_sed}, the total continuum flux is higher than the fit for 0.85 mm 
and 0.45, because of their lower resolutions. However the 1.3 cm continuum data 
can resolve the G35.58-0.03 into two sources (see Figure \ref{fig_nh3_mom0}),
which leads to a low flux. So, for the 1.3 cm continuum, we used 
``IMSTAT'' (AIPS task) to obtain a total flux of $\sim$255 mJy and a peak flux 
of $\sim$80.29$\,\mjyb$ within a region of $\sim1\dotsec5\times1\dotsec5$.
If doing ``JMFIT'' (AIPS task) with one Gaussian component, we will just obtain a total flux with $184.4\pm2.7$ mJy 
and the peak flux with $68.8\pm0.7\,\mjyb$ at 1.3 cm. Due to the data contamination, 
the SED fitting is rough relatively.

\subsection{Hypercompact \HII Region} \label{sect:HII}

The HC \HII region G35.58-0.03 is unresolved in the SMA 1.3 mm continuum
observation in Figure \ref{fig_br}, while it presents an elongated
distribution with PA=23$\dotdeg$7 in the VLA 1.3 cm continuum
observation in Figure \ref{fig_nh3_mom0}. The elongated distribution 
has a direction consistent with an outflow (see Section \ref{sect:outflow}), suggesting that 
there exists interaction between the ionized and molecular gas.
We derive an intrinsic
size $\theta_s=\sqrt{0\dotsec655\times0\dotsec192}\approx0\dotsec354$ for the
HC \HII core from the deconvolved beam size of the 1.3 cm
continuum observation. 
For optically thin ionized gas and local
thermodynamic equilibrium (LTE) condition, one can derive 
the electron temperature $T_e^*$ based on the 1.3 mm continuum flux density and the 
H30$\alpha$ line intensity \citep{gord2002,wils2009,shih2010}:
\begin{eqnarray}
T_e^*&=&\Bigg[\left(\frac{6985}{\alpha(\nu,T_e)}\right)
     \left(\frac{\Delta V_{\rm H30\alpha}}{\rm km~s^{-1}}\right)^{-1}  
     \left(\frac{S^{\rm gas}_{\rm 1.3\,mm}}{S_{\rm H30\alpha}}\right)
     \times \nonumber \\    &&  
     \left(\frac{\nu}{\rm GHz}\right)^{1.1}
    \left(1+\frac{N({\rm He^+})}{N({\rm H^+})}\right)^{-1}\Bigg]^{0.87},
\end{eqnarray}
where $\alpha(\nu,T_e) \sim 1$ is a slowly varying function tabulated by
\citet{mezg1967}, and $\frac{N({\rm He^+})}{N({\rm H^+})} \sim 0.096$ was adopted
\citep{mehr1994}. $S^{\rm gas}_{\rm 1.3\,mm} = 0.811\times55\%~\jyb$ (upper limit from Section \ref{sect:sed})
and $S_{\rm H30\alpha} = 1.34~\jyb$ are the peak flux density. Finally we get an electron
temperature $T_e^*=5500$\,K. Due to the uncertainty in the estimate of the dust 
contribution to $S_{\rm 1.3\,mm}$, the error on the temperature might be high.
This temperature is a lower limit due to the uncertain SED fitting flux density, and
the 1.3 mm data in relatively low resolution tracing also extended envelope material.

\begin{figure}
\figurenum{9}
\begin{center}
\includegraphics[angle=-90,scale=0.35]{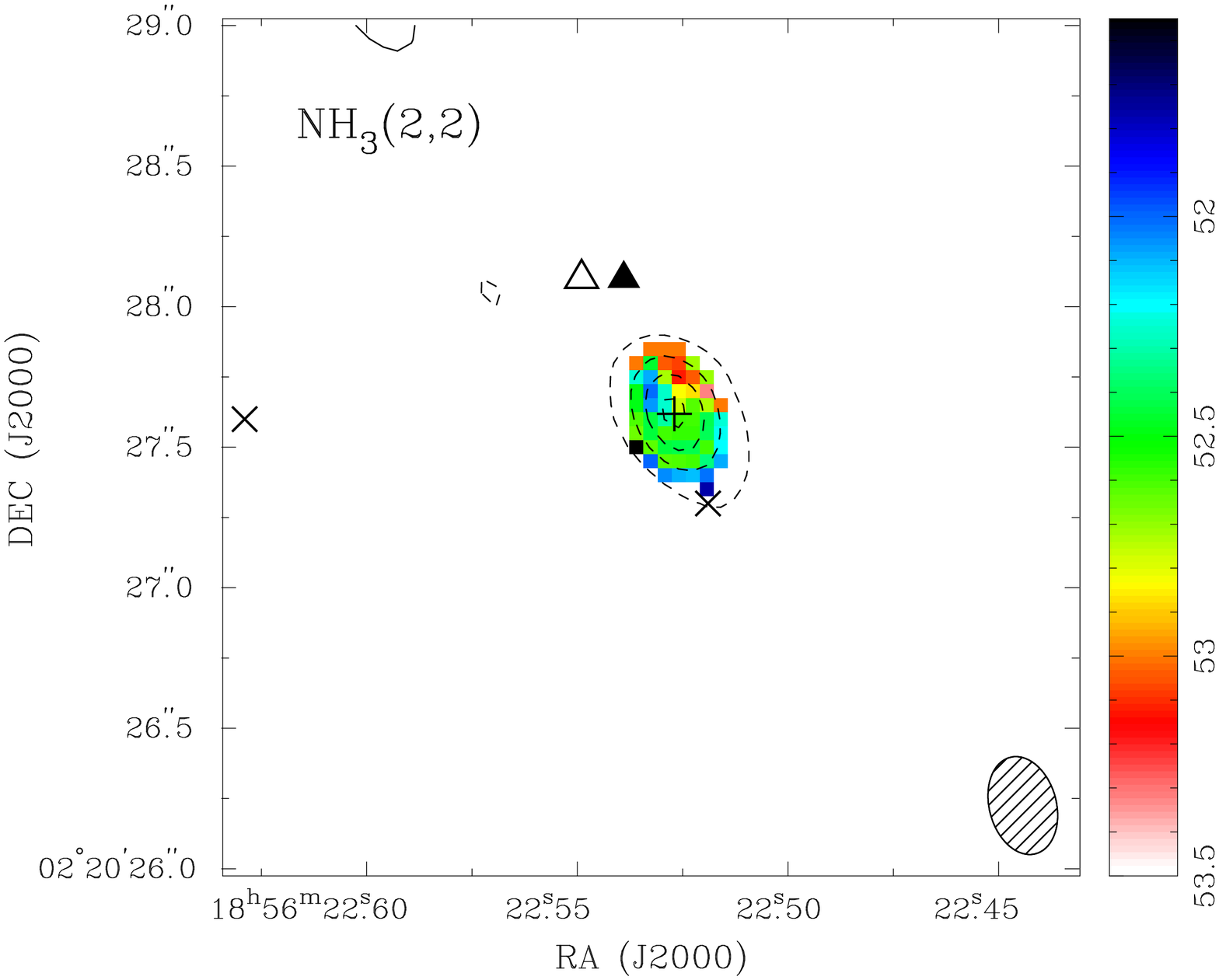}
\includegraphics[angle=-90,scale=0.35]{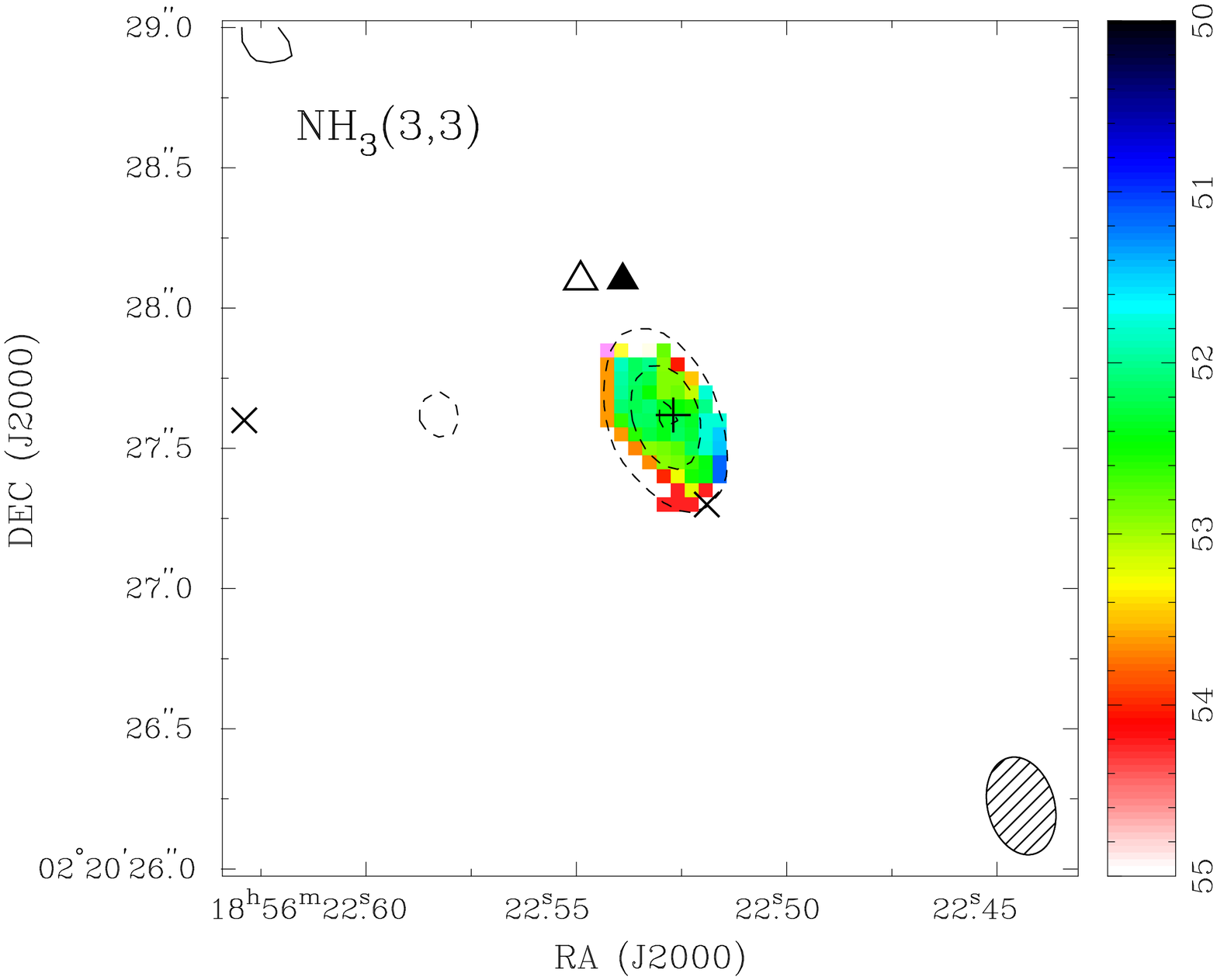}
\caption{  Velocity-integrated contours of NH$_3$ (2, 2) and (3, 3)
superimposed on the intensity-weighted mean velocity (moment 1, {\it
color scale}) maps. Contours and symbols are the same as in Figure
\ref{fig_nh3_mom0}. } \label{fig_nh3_mom1}
\end{center}
\end{figure}

The emission measure (EM) is an important parameter for the HC \HII region. 
We derive the EM parameter from equation in \citet{wils2009} and \citet{shih2010} with
\begin{eqnarray}
{\rm EM} &=& 7.1~{\rm pc~cm^{-6}} 
    \left(\frac{S_{L}}{\rm Jy}\right)
    \left(\frac{\lambda}{\rm mm}\right)
    \left(\frac{T_e}{\rm K}\right)^{1.5}  \times \nonumber \\  &&  
    \left(\frac{\Delta V}{\rm km~s^{-1}}\right)  
    \left(\frac{\theta_s}{\rm arcsec}\right)^{-2}. \label{eq:rrlEM}
\end{eqnarray}
Assuming $T_e\approx T_e^*$, the electron temperature is estimated by $T_e \sim$ 
5500 K for the HC \HII core. The intrinsic
size is $\theta_s\approx0\dotsec354$ for the \HII region. 
For the H30$\alpha$ line, the peak line intensity is $S_L$ = 1.43 Jy, 
and the FWHM is $\Delta V = 43.16~\kms$. The observing wavelength is $\lambda$ = 1.3 mm. 
Based on the parameters above, we obtain an EM $= 1.9\times10^{9}$ pc cm$^{-6}$ for the HC \HII region. 
The corresponding volume electron density is estimated by
$n_e= 3.3\times10^{5}$\,cm$^{-3}$ from EM = $n_e^2Lf_V$ ($L$ and $f_V$ is the path 
length and volume filling factor, respectively), assuming the HC \HII core size is
$L$ = $10.2\times tan(0\dotsec354)$ kpc = 3714 AU, and $f_V$ = 1.
The continuum optical depth $\tau_c$ can be obtained from the equation in \citet{mezg1967} with
     \begin{eqnarray}
       \tau_c &=& 0.08235~\alpha(\nu, T_e) 
                   \left(\frac{\nu}{{\rm GHz}}\right)^{-2.1}
                   \left(\frac{T_e}{{\rm K}}\right)^{-1.35}  \times \nonumber \\    && 
                   \left(\frac{\rm EM}{{\rm pc~cm^{-6}}}\right).
                                   \label{eq:tau_c}
     \end{eqnarray}
$\tau_c\approx0.015$ is finally derived, assumed $\alpha(\nu, T_e)\sim1$.
However, the same formula (\ref{eq:tau_c}) gives an optical depth of about
1.8, if the frequency of the 1.3 cm continuum is used.
The derived parameters above indicate that the H30$\alpha$ line is optically thin, and 
traces a very dense region ($<$ 3714 AU) with high temperature. In addition, it is necessary 
to further use RRLs as tracers to study the regions with high density and temperature.

We employ the Lyman continuum photon number ($N_L$) and the excitation parameter
($U$) to infer the corresponding star type. The corresponding equations 
are expressed \citep{rubi1968,pana1973,mezg1974,mats1976}:
     \begin{eqnarray}
       N_{L} &=& 4.761 \times 10^{48} {\rm s^{-1}}
                 \alpha\left(\nu, T_e\right)^{-1}
                 \left(\frac{\nu}{\rm GHz}\right)^{0.1}
                 \left(\frac{D}{\rm kpc}\right)^{2}  \times \nonumber\\   &&    
                 \left(\frac{S^{\rm gas}_{\nu}}{\rm Jy}\right)
                 \left(\frac{T_e}{\rm K}\right)^{-0.45},
                 \label{eq:NL} \\
       U &=& 2.706\times10^{-16} {\rm pc~cm^{-2}}
                                   \left(\frac{T_e}{\rm K}\right)^{4/15}
                                   \left(\frac{N_L}{\rm s^{-1}}\right)^{1/3},
     \end{eqnarray}
where $D$ = 10.2 kpc is the source distance from the Sun, and $\alpha(\nu,
T_e)\sim1$ is used again. The upper limit of the continuum free-free emission 
flux is $S^{\rm gas}_{\nu} = 1.056\times55\%$ Jy at 1.3
mm. Based on our previous derived parameters assuming optically thin and
LTE conditions, we obtained two upper limits of $N_L$ = 1.0$\times$10$^{49}$ s$^{-1}$ and
$U$ = 58.0\,pc~cm$^{-2}$. The inferred Lyman continuum photon number requires
a massive star equivalent to a zero-age main-sequence star of type
O6.5 located inside the HC \HII region G35.58-0.03 \citep{pana1973}.

\subsection{Radio Recombination Lines H30$\alpha$ and H38$\beta$} \label{sect:rrl}

\begin{figure}
\figurenum{10}
\begin{center}
\includegraphics[angle=-90,scale=0.35]{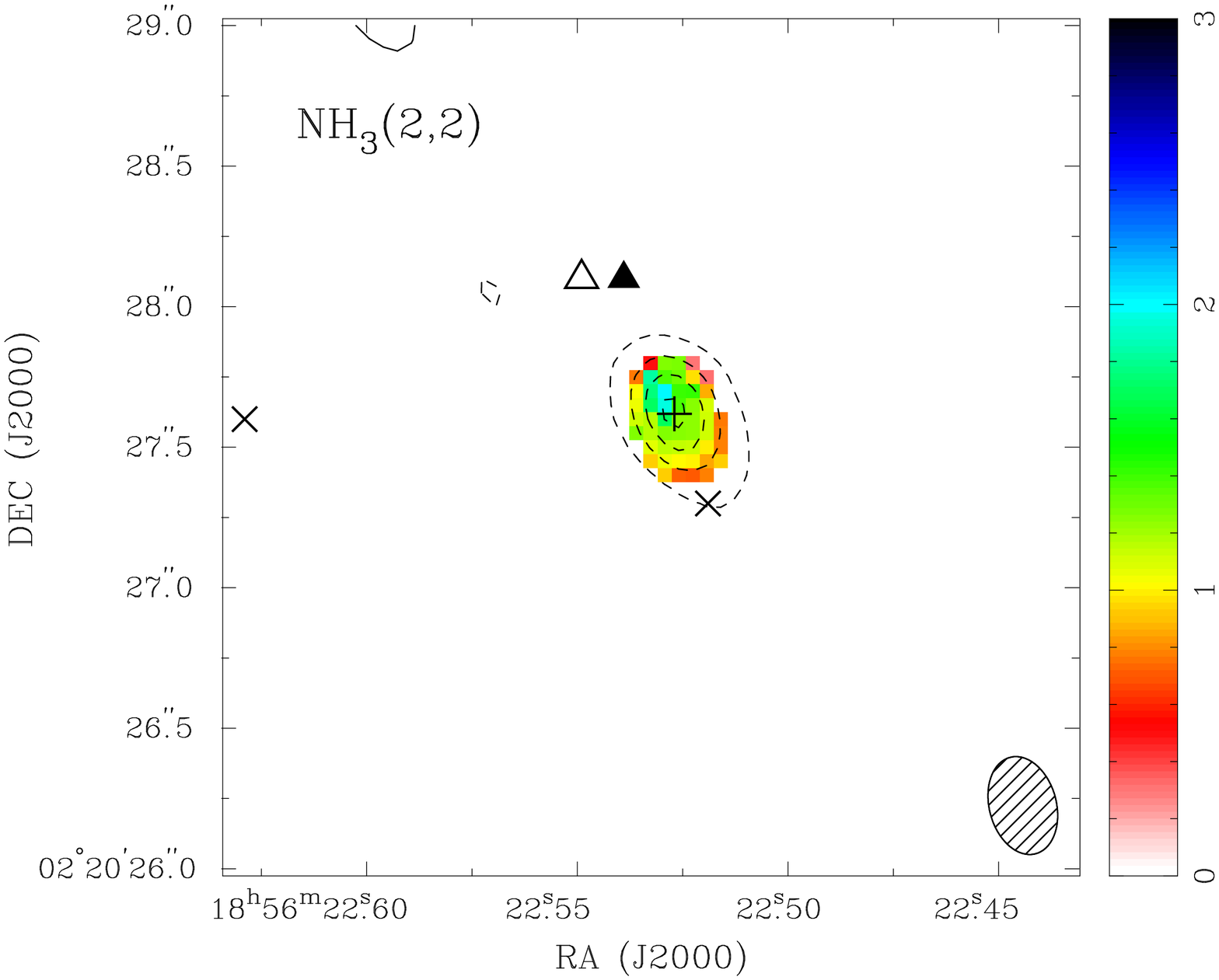}
\includegraphics[angle=-90,scale=0.35]{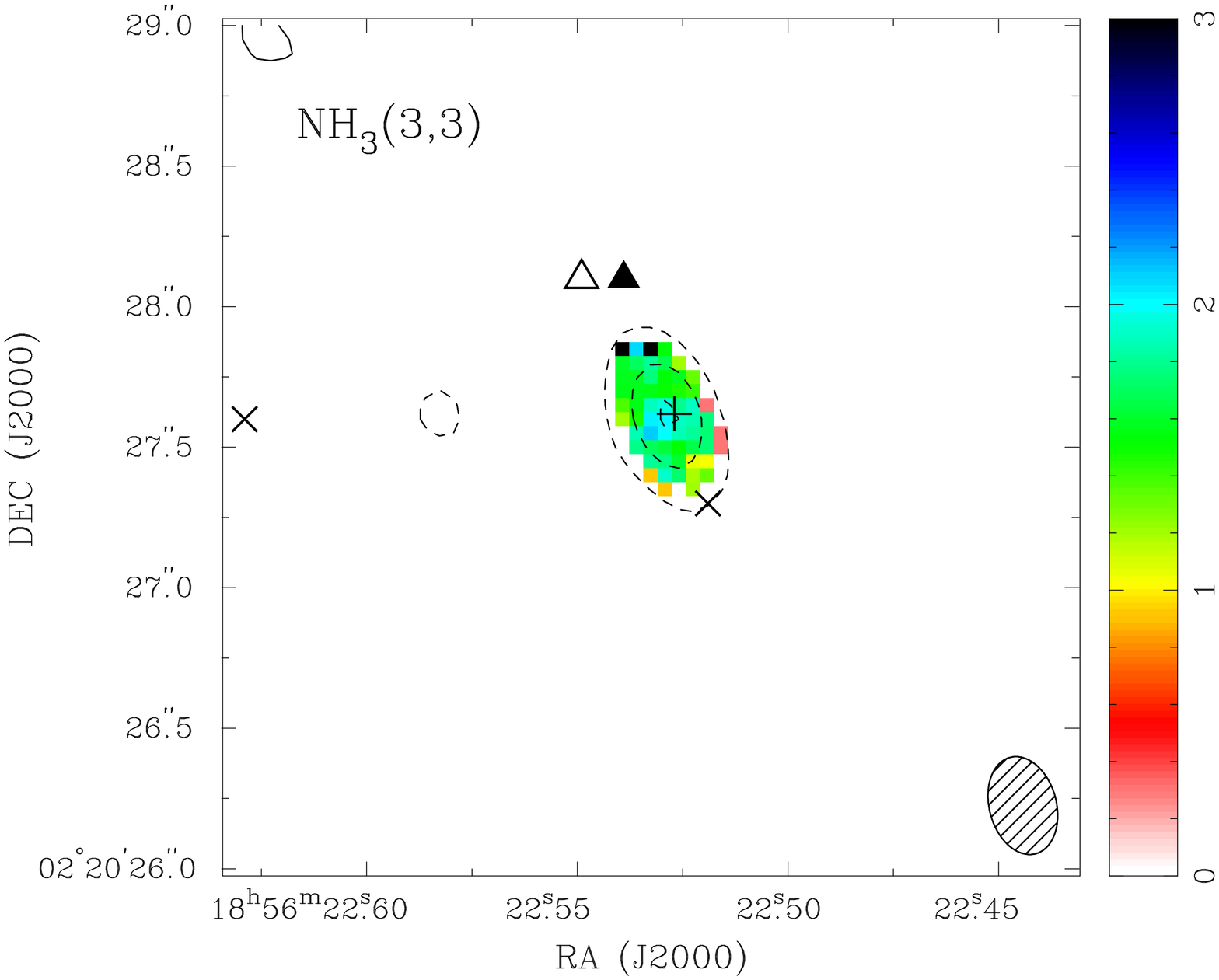}
\caption{ Velocity-integrated contours of NH$_3$ (2, 2) and (3, 3)
superimposed on the velocity dispersion (moment 2, {\it color
scale}) with respect to moment-1 velocity maps. Contours and symbols
are the same as in Figure \ref{fig_nh3_mom0}. } \label{fig_nh3_mom2}
\end{center}
\end{figure}

The observed line width, $\Delta V$, of the H30$\alpha$ and H38$\beta$ lines is produced
by a combination of pressure broadening, $\Delta V_P$, and Doppler broadening, $\Delta V_D$. 
The Doppler broadening, $\Delta V_D$, includes thermal
broadening, $V_{ther}$, which is due to the thermal motion of the
particles, and the dynamical and/or turbulent broadening, $\Delta V_{dyn}$, which is
due to infall, outflow, and/or rotation movements. In this work, the H38$\beta$ 
line ($S_{{\rm H}38\beta}$ $\approx$ 0.26 $\jyb$) is much weaker than 
the H30$\alpha$ line ($S_{{\rm H}30\alpha}$ $\approx$ 1.34 $\jyb$), but both 
have almost the same line width ($\Delta V_{{\rm H}30\alpha}$ $\approx$ 43.2\,km~s$^{-1}$, 
and $\Delta V_{{\rm H}38\beta}$ $\approx$ 43.2\,km~s$^{-1}$).

\begin{figure*}
\figurenum{11}
\begin{center}
\includegraphics[angle=0,scale=0.99]{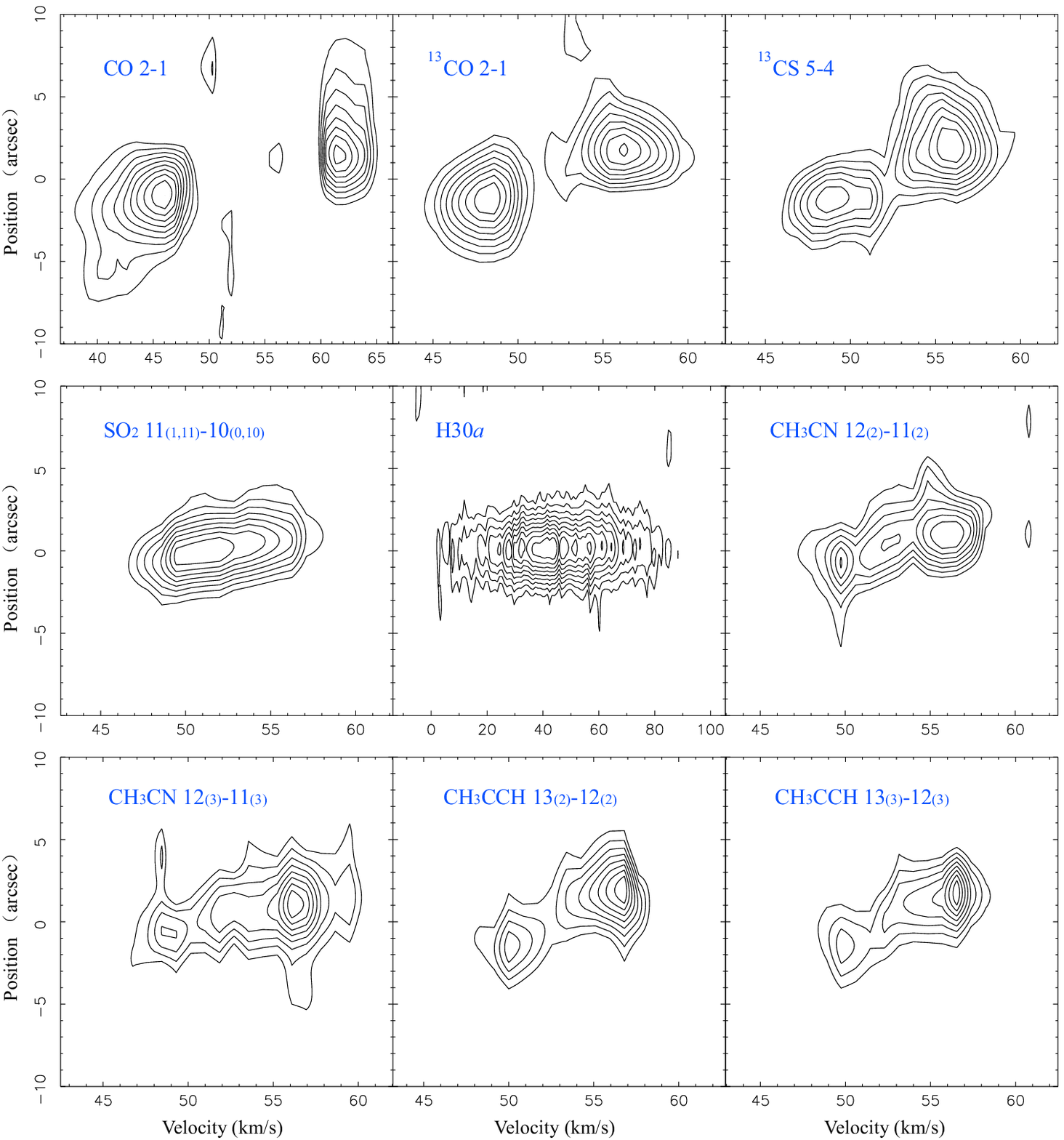}
\end{center}
\caption{ PV diagrams of 1.3 mm observations with P.A. = 90$^\circ$
at the position of the 1.3 mm continuum peak. Contours of CO (2-1)
are 2.30, 4.60, 6.89, 9.19, 11.5, 13.8, 16.1, 18.4, 20.7 K; Contours
of $^{13}$CO (2-1) are 5.47, 8.20, 10.9, 13.7, 16.4, 19.1, 21.9,
24.6 K; Contours of $^{13}$CS (5-4) are 0.616, 0.924, 1.23, 1.54,
1.85, 2.16, 2.47, 2.77 K; Contours of SO$_2$ 11(1,11)-10(0,10) are
0.839, 1.26, 1.68, 2.10, 2.52, 2.94, 3.36, 3.78 K; Contours of
H30$\alpha$ are 0.317, 0.634, 0.951, 1.27, 1.58, 1.90, 2.22, 2.54,
2.85 K; Contours of CH$_3$CN 12(2)-11(2) are 0.361, 0.481, 0.601,
0.721, 0.841, 0.962, 1.08 K; Contours of CH$_3$CN 12(3)-11(3) are
0.304, 0.456, 0.608, 0.760, 0.912, 1.06, 1.22, 1.37 K; Contours of
CH$_3$CCH 13(2)-12(3) are 0.423, 0.635, 0.847, 1.06, 1.27, 1.48,
1.69, 1.90 K; Contours of CH$_3$CCH 13(3)-12(3) are 0.501, 0.751,
1.00, 1.25, 1.50, 1.75, 2.00, 2.25 K.} \label{fig_pv}
\end{figure*}

The broadening mechanisms can be described by the following equations \citep{gord2002,shih2010}: 
     \begin{eqnarray}
       \Delta V&=& \sqrt{\Delta V_P^2+\Delta V_{D}^2},                  \\
       \Delta V_D&=&\sqrt{\Delta V_{dyn}^2+\Delta V_{ther}^2} \nonumber \\
       &=&\sqrt{\left(\frac{\Delta V_{dyn}}{\rm km~s^{-1}}\right)^{2}
       +0.0458\left(\frac{T_e}{\rm K}\right)}~{\rm km~s^{-1}},\\
       \Delta V_P&=&
                  3.74\times10^{-14}~n^{4.4}~{\rm km~s^{-1}}
                   \left(\frac{n_e}{\rm cm^{-3}}\right) \times \nonumber \\                 &&
                  \left(\frac{\lambda}{\rm mm}\right)
                 \left(\frac{T_e}{\rm K}\right)^{-0.1}, \label{eq:V_P}
     \end{eqnarray}
where $n$ = 30 and 38, respectively, is the principal quantum number 
of the H30$\alpha$ and H38$\beta$ transitions. 
The broadening of the lines is derived using a local electron density
($n_e\approx3.3\times$10$^5$\,cm$^{-3}$) and electron temperature
($T_e\approx5500$\,K) for the HC \HII region. The
thermal broadening is $\Delta V_{ther}\sim$15.87\,km~s$^{-1}$
for $T_e=5500$\,K. Due to the relatively small principal quantum number, the 
pressure broadening ($\Delta V_P$(H30$\alpha$) = 0.02\,km~s$^{-1}$, and 
$\Delta V_P$(H38$\beta$) = 0.06\,km~s$^{-1}$) contributes little to the 
total line width. The broadening contribution from the dynamical and/or turbulent broadening motions
is $\Delta V_{dyn}$(H30$\alpha$) $\approx$~40.14\,km~s$^{-1}$ and 
$\Delta V_{dyn}$(H38$\beta$) $\approx$~40.16\,km~s$^{-1}$ for
the H30$\alpha$ and H38$\beta$ lines, respectively. In addition, it also can be
seen from the moment-1 map (Figure \ref{fig_sma_mom1}) and the PV
diagram (Figure \ref{fig_pv}) that the H30$\alpha$ line shows
evidence of blueshifted and redshifted wings, which probably suggests
infall, outflow, and/or rotation. Therefore, the H30$\alpha$ line
traces the high-temperature ionized gas, which is participating
in dynamical movements.

\subsection{Rotational Temperature}\label{sect:rota}

\begin{figure}
\figurenum{13}
\begin{center}
\includegraphics[angle=0,scale=0.62]{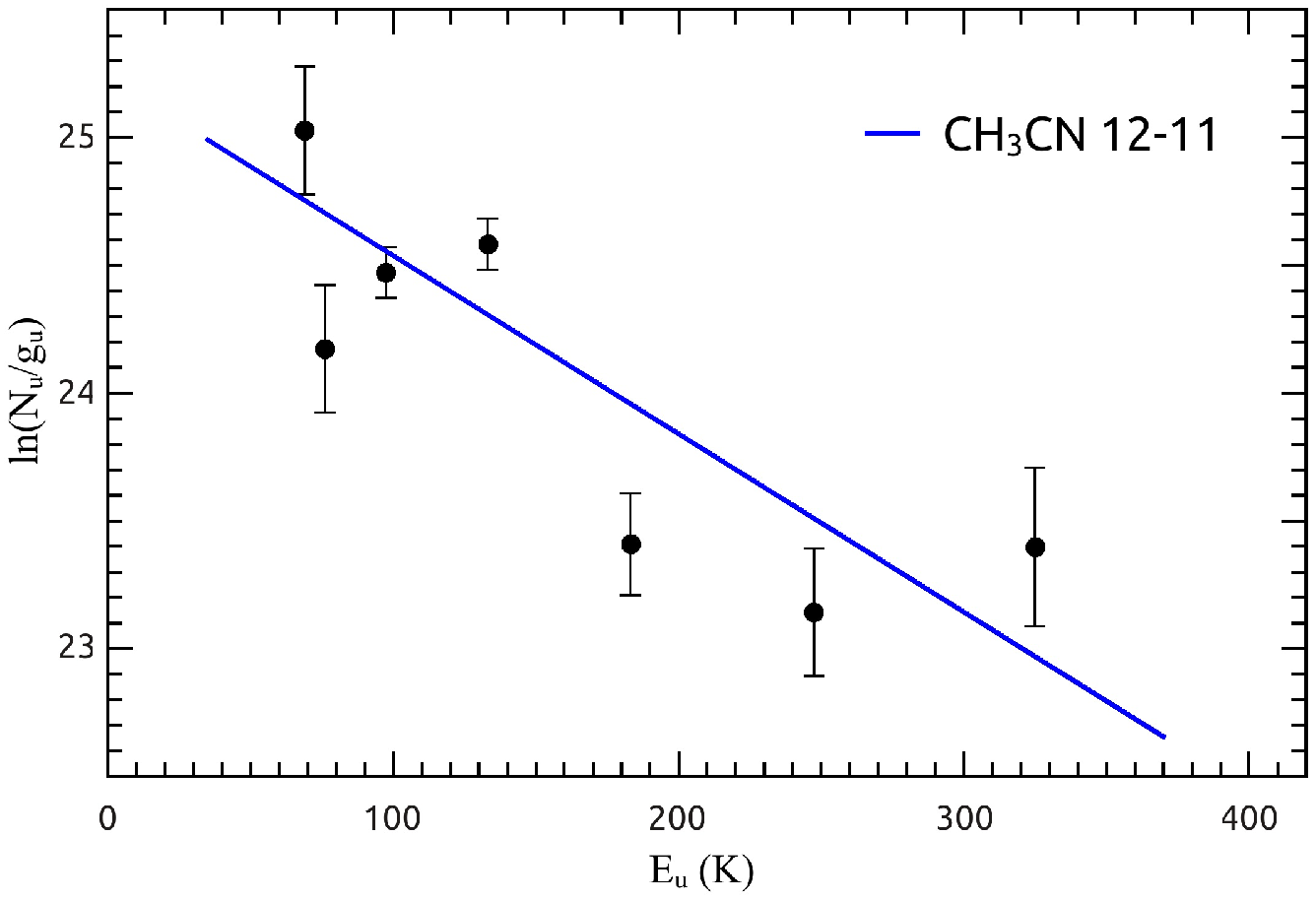}
\includegraphics[angle=0,scale=0.62]{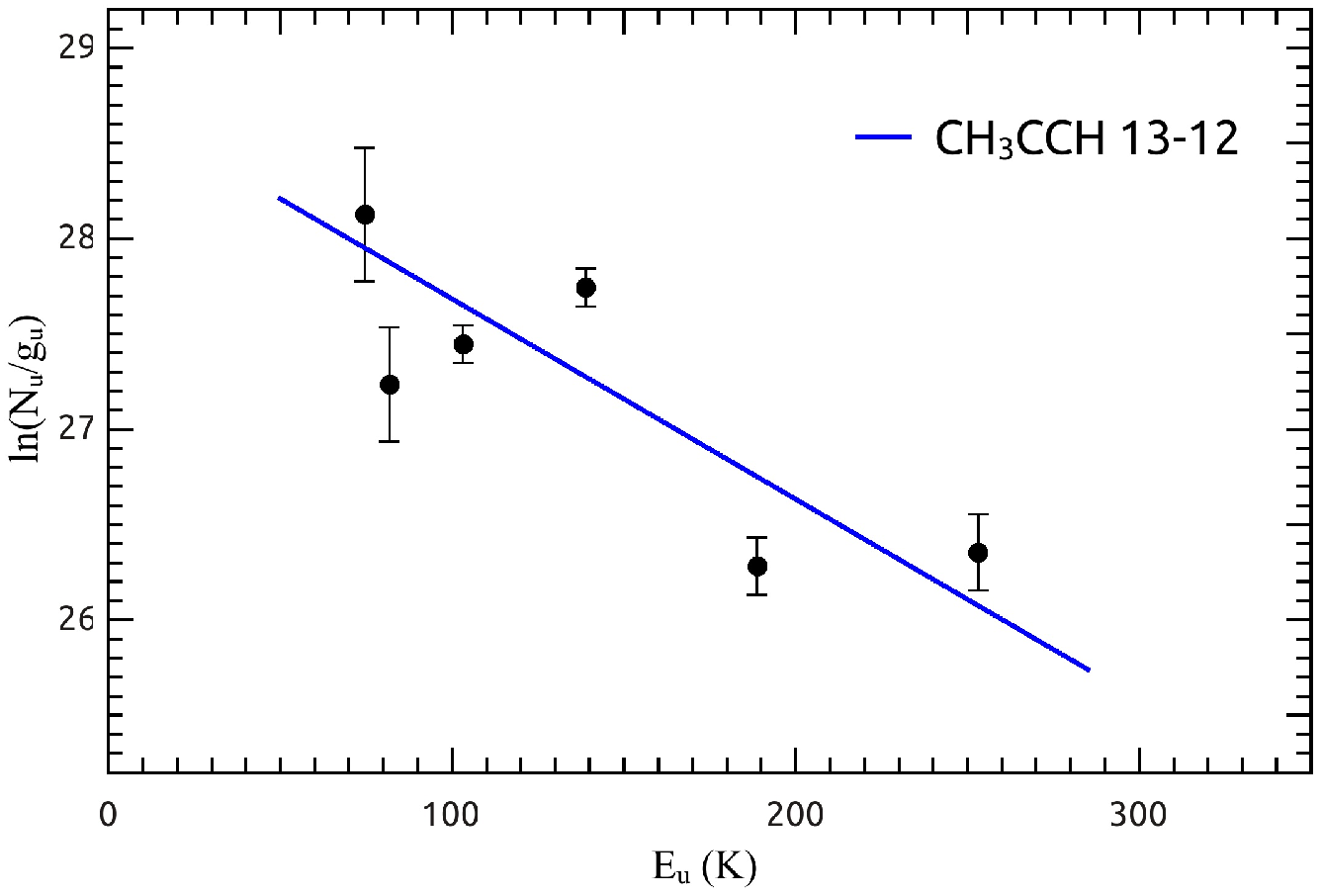}
\caption{ Rotational temperature diagrams for the CH$_3$CN and CH$_3$CCH. The filled 
circles with vertical error bars are for the observed transitions. The linear least-square 
fittings are shown as the solid blue line. The rotational temperatures derived from the 
diagram method are T$_{\rm rot}$ = 143 $\pm$ 20 K for CH$_3$CN and T$_{\rm rot}$ = 95 $\pm$ 
17 K for CH$_3$CCH, respectively. The relevant parameters are listed in Table \ref{tab_rota}.} 
\label{fig_rotat}
\end{center}
\end{figure}

The rotational temperature diagram (RTD) method can be used to
estimate the rotational temperature, if multiple transitions with
different upper level energies are observed simultaneously
\citep{gold1999,aray2005}. It may give an average excitation temperature of the specific
molecule even in non-LTE cases \citep{qins2010}.
The corresponding parameters for the rotational temperature fitting are listed in Table \ref{tab_rota}.
Due to the blending of CH$_3$CN 12(0)-11(0) and 12(1)-11(1), and CH$_3$CCH 13(0)-12(0) and 13(1)-12(1), 
we used their 5$\sigma$ as error in the fit. Following the RTD method from
\citet{qins2010}, we get rotational temperatures of $\sim$ 143 $\pm$ 20 K for
CH$_3$CN (12-11), and $\sim$ 95 $\pm$ 17 K for CH$_3$CCH (13-12) in Figure \ref{fig_rotat}. 
The high rotational temperature suggests that the CH$_3$CN and CH$_3$CCH transitions originate from warm
gas environment. The derived the rotational temperature (143 and 95 K) of the two molecules from the HC \HII region is equivalent to the canonical value (100 K) from the hot core stage.

\subsection{Infall, Outflow, and/or Rotation} 

\subsubsection{Infall}\label{sect:infall}

In Figure \ref{fig_spectra}, the CO (2-1), $^{13}$CO
(2-1), $^{13}$CS (5-4), OCS (19-18), $^{34}$SO$_2$
22(2,20)-22(1,21), and $^{34}$SO$_2$ 13(2,12)-13(1,13) lines show
double-peak profiles with blueshifted peaks stronger than redshifted
peaks. To some extent, flux is missing in the line center of the CO (2-1) and $^{13}$CO
(2-1) lines. However, the infall evidence can be determined using the red and blue shifted peaks alone. 
The red peaks of both SO$_2$ 11(1,11)-10(0,10) (in Figure
\ref{fig_spectra}) and H30$\alpha$ (in Figure \ref{fig_ch3cn}) are also
absorbed, compared to the blue ones. Also the CO (2-1) spectrum shows a
prominent blue shifted profile. In Figure \ref{fig_pv}, the PV
diagrams of 1.3 mm lines except for H30$\alpha$ show obvious
velocity gradients. These spectral features indicate an infalling
circumstellar envelope surrounding the compact core of the HC \HII
region G35.58-0.03.

The rapidly infalling envelope can provide the force to prevent the
expansion of the \HII region. The infall velocity is $V_{\rm in}$ = $\sim$1.5
$\kms$ by comparing the systemic velocity $\sim$52.5 $\kms$ with the
velocity of the redshifted absorbing dip ($\sim$54 $\kms$) in CO
(2-1), OCS (19-18), $^{34}$SO$_2$ 22(2,20)-22(1,21), and
$^{34}$SO$_2$ 13(2,12)-13(1,13). Because of the CO lines being optically thick, and the missing flux due to a
lack of short spacing data, we use the 1.3 mm continuum to
derive the column density and mass of the \HII region envelope.  The mean dust temperature can be
estimated as a lower limit of $\sim$ 95 $\pm$ 17 K from the rotational
temperature of CH$_3$CN and CH$_3$CCH in Section \ref{sect:rota}. Assuming an average grain radius of 0.1
$\mu$m and grain density of 3 g cm$^{-3}$ and a gas to dust ratio of
100, the hydrogen column density is given by the formula
\citep{lisd1991}
     \begin{eqnarray}
       N_{\rm{H_{2}}} &=& 8.1 \times 10^{17} {\rm cm^{-2}}
               \frac{e^{h\nu/kT}-1}{\rm Q(\nu) \Omega}
                \left(\frac{S^{\rm dust}_{\nu}}{\rm Jy}\right)
                  \left(\frac{\nu}{\rm GHz}\right)^{-3},
                   \label{eq:NL}
     \end{eqnarray}
where T $\sim$ 95 K is the mean dust temperature,
Q($\nu$)$\,\sim\,2\times10^{-5}$ is grain emissivity at frequency 231
GHz, $S^{\rm dust}_{\nu}=1.056\times75\%$ Jy is the warm dust emission of the continuum at
1.3 mm peak, and $\Omega$ is the beam solid angle. The derived hydrogen
column density is $N_{\rm{H_{2}}}$ = $1.7\times10^{24}$ cm$^{-2}$,
so that the volume density is $n_{\rm{H_{2}}}$ = $3.5\times10^{6}$
cm$^{-3}$. In addition, the total envelope mass in a beam size can be estimated
from the 1.3 mm continuum using the formula \citep{lisd1991}
     \begin{eqnarray}
       M_{\rm{H_{2}}} &=& 1.3 \times 10^{4} {\rm M_{\odot}}
               \frac{e^{h\nu/kT}-1}{\rm Q(\nu)}
                \left(\frac{S^{\rm dust}_{\nu}}{\rm Jy}\right) \times \nonumber \\  && 
                  \left(\frac{\nu}{\rm GHz}\right)^{-3}
                  \left(\frac{D}{\rm kpc}\right)^{2}.
                   \label{eq:M_env}
     \end{eqnarray}
Assuming the molecular mass ratio
$M_{\rm{env}}/M_{\rm{H_{2}}}=1.36$, we obtain a total envelope mass
$M_{\rm{env}}=538~{\rm M_{\odot}}$ as an upper limit. The infall rate of the envelope
material can be estimated by
\begin{eqnarray}
 \dot{M}_{\rm in}
              &=& 2.1\times10^{-11} {\rm M_{\sun}~yr^{-1}}
              \left(\frac{V_{\rm in}}{\rm km~s^{-1}}\right)
                \left(\frac{n_{\rm H_2}}{\rm cm^{-3}}\right) \times \nonumber \\   && 
                \left(\frac{D}{\rm kpc}\right)^2 
                \left(\frac{\theta_{\rm in}}{\rm arcsec}\right)^2,
         \label{eq:dMdt}
\end{eqnarray}
where $\theta_{\rm
in}=\sqrt{2\dotsec238\times1\dotsec285}\approx1\dotsec696$ (from the
deconvolved beam size of the 1.3 mm continuum observation) is the
diameter of the infalling region, $D$ is the distance to the source,
and $V_{\rm in} \sim 1.5\,\kms$ is the infall velocity. A lower limit of the infall
rate of 0.033 M$_{\sun}$~yr$^{-1}$ is inferred with the derived
parameters. The momentum of the infalling gas per year is $\sim 0.050$ M$_{\sun}\,\kms$.

A high accretion rate is necessary to form an O star, however,
\cite{walm1995} reports a high ``critical accretion rate'' to ``choke off'' an 
\HII region formation. In this work, we obtain an infall rate of 0.033 M$_{\sun}$~yr$^{-1}$,
which is high enough to quench any \HII region and to overcome the radiation 
pressure from central star. The observed lines at 1.3 cm do not show any
infalling features (see Section \ref{sect:ammonia}). This suggests that the accretion in the inner
part might already be halted.

\subsubsection{Outflow} 
\label{sect:outflow}

Figure \ref{fig_spectra} shows the CO (2-1) spectrum with very broad
wings between $\sim$38 and $\sim$65 $\kms$. High-velocity gas can
be easily identified in the PV diagrams of Figure \ref{fig_pv} and is exhibited in Figure \ref{fig_br}. The
Gaussian fitting FWHM of CO (2-1) spectrum is $\sim$13.53 $\kms$. In
all spectra shown in Figures \ref{fig_spectra}, \ref{fig_ch3cn}, and
\ref{fig_nh3_spec}, the narrowest two spectra are CH$_3$CCH
13(4)-12(4) with FWHM $\approx$ 3.99 $\kms$, and NH${_3}$ (3, 3)
with FWHM $\approx$ 4.35 $\kms$. Figure \ref{fig_br} exhibits a
bipolar outflow in the tracers CO (2-1), $^{13}$CO (2-1), $^{13}$CS
(5-4), and SO$_2$ 11(1,11)-10(0,10). The two lobes can also be
identified from the intensity-weighted mean velocity in
Figure \ref{fig_sma_mom1}. Especially, there is one consistent direction 
of the velocity gradient between H30$\alpha$ and other molecular gas seen from
Figure \ref{fig_sma_mom1}D with \ref{fig_sma_mom1}A, \ref{fig_sma_mom1}B, and \ref{fig_sma_mom1}C,
indicating that the ionized outflow is acting along with the molecular 
bipolar outflow. This is evidence for an ionized outflow driving a molecular
outflow \citep{klaa2013}. The H30$\alpha$ line has a large velocity dispersion at
the position of 1.3 cm continuum peak in Figure \ref{fig_sma_mom2},
suggesting its driving source position. 1.3 mm, 1.3 cm, and 3.6 cm
continuum locations are well correlated with the outflow. Water and
OH masers were also found near the position of the continuum peaks.
This shows that the outflow is very active. There are small differences in the outflow directions seen
in Figure \ref{fig_br}. We argue that $^{13}$CS (5-4)
and SO$_2$ 11(1,11)-10(0,10) outflows are possibly contaminated
with infall and rotation. Furthermore, the CO (2-1) and $^{13}$CO (2-1) outflows,
showing also lobes to the north (see Figure \ref{fig_br}) in the direction
perpendicular to the bipolar outflow, maybe also be blended with infall
and rotation.

\begin{center}
\begin{deluxetable*}{lccccc}
\tabletypesize{\scriptsize} \tablecaption{Rotational temperature diagram parameters\tablenotemark{a} \tablenotemark{b}} \tablewidth{0pt} \tablehead{
\colhead{Molecule} & \colhead{Transition} & \colhead{$S\mu^2$}  &  \colhead{$E_u$}  &  \colhead{Flux}   &  \colhead{Intensity}  \\
	&		&	debye$^2$  	&	K	&	Jy/beam$\cdot$km/s	&	Jy/beam	\\
(1)  &   (2)   &  (3)  &  (4) &  (5) & (6)} \startdata
CH$_{3}$CN   &   12(0)-11(0)	&	183.74	&	69	&	6.39(0.39)	&	0.67(0.09)	\\
CH$_{3}$CN   &   12(1)-11(1)	&	182.46	&	76	&	4.02(0.59)	&	0.36(0.09)	\\
CH$_{3}$CN   &   12(2)-11(2)	&	178.64	&	97	&	4.12(0.35)	&	0.41(0.06)	\\
CH$_{3}$CN   &   12(3)-11(3)	&	172.26	&	133	&	4.44(0.44)	&	0.45(0.06)	\\
CH$_{3}$CN   &   12(4)-11(4)	&	163.32	&	183	&	1.30(0.50)	&	0.18(0.06)	\\
CH$_{3}$CN   &   12(5)-11(5)	&	151.84	&	247	&	0.93(0.36)	&	0.17(0.09)	\\
CH$_{3}$CN   &   12(6)-11(6)	&	137.80	&	325	&	1.07(0.33)	&	0.16(0.09)	\\
CH$_{3}$CCH  &   13(0)-12(0)	&	7.92	&	75	&	5.96(0.35)	&	0.63(0.09)	\\
CH$_{3}$CCH  &   13(1)-12(1)	&	7.87	&	82	&	3.34(0.35)	&	0.46(0.09)	\\
CH$_{3}$CCH  &   13(2)-12(2)	&	7.73	&	103	&	3.56(0.37)	&	0.40(0.08)	\\
CH$_{3}$CCH  &   13(3)-12(3)	&	7.50	&	139	&	4.65(0.38)	&	0.50(0.07)	\\
CH$_{3}$CCH  &   13(4)-12(4)	&	7.17	&	189	&	1.03(0.33)	&	0.33(0.08)	\\
CH$_{3}$CCH  &   13(5)-12(5)	&	6.75	&	253	&	1.04(0.35)	&	0.21(0.08)	\\
\hline
\enddata
\tablenotetext{a}{Blend of CH$_{3}$CN at 12(0)-11(0) and 12(1)-11(1).}
\tablenotetext{b}{Blend of CH$_{3}$CCH at 13(0)-12(0) and 13(1)-12(1).}
\label{tab_rota}
\end{deluxetable*}
\end{center}

Many studies show that low-velocity molecular gas toward the core
is usually optically thick \citep{gold1984,snel1984}. Under
conditions of local thermodynamic equilibrium (LTE), we assume that
both the blueshifted and redshifted lobes are optically thin for
$^{13}$CO (2-1). The relation between opacities and the ratio of
$^{12}$CO (2-1) to $^{13}$CO (2-1) main-beam brightness temperature
\citep{myer1983} is
     \begin{eqnarray}
\frac{T_{\rm MB}(^{12}{\rm CO})}{T_{\rm MB}({\rm
^{13}CO})}&=&\frac{1-e^{-\tau^{12}}}{1-e^{-\tau^{13}}},
     \end{eqnarray}
where we assume $\tau^{12} = 89\tau^{13}$ \citep{lang1980,gard1991}.
Furthermore, the excitation temperature $T_{ex}$ is derived from the
equation of radiative transfer
\begin{equation}
\left\{ \begin{aligned}
         T_{\rm MB}&=f[J(T_{ex})-J(T_{bg})][1-e^{-\tau}] & \\
         J(T)&=T_{0}/[e^{T_{0}/T}-1] &
\end{aligned}, \right.
\end{equation}
where $f$ is the beam filling factor, $T_{bg}$ = 2.7 K is background
temperature, and $T_{0}$ = $h\nu/k$ for the transition of $^{13}$CO
(2-1) \citep{wong2008}. We then obtain the molecular $^{13}$CO
(2-1) column density $N(^{13}$CO) from the relation
\begin{equation}
    N(^{13}{\rm CO})=1.51\times10^{14}{\rm cm^{-2}}\frac{e^{5.3/T_{\rm
ex}}}{1-e^{-10.6/T_{\rm{ex}}}}\int T_{\rm MB}{\rm d}v.
\end{equation}
If we assume that the [H$_{2}$/$^{13}$CO] abundance ratio is
$8.9\times10^{5}$ \citep{gard1991}, the molecular hydrogen column density,
$N( \rm H_{2})$, can be calculated. The total mass of gas in the
synthesized beam can be calculated from equation
     \begin{equation}
      M_{\rm out}=4.57\times10^{-19}{\rm M_{\odot}}
      \left( \frac{\theta_{\rm beam}}{\rm arcsec} \right)^2
     \left( \frac{N(^{13}{\rm CO})}{\rm cm^{-2}} \right)
       \left( \frac{D}{\rm kpc} \right)^2,
     \end{equation}
where we assume that the mean atomic weight of the gas is $\mu_{g}$
= 1.36, $\theta_{\rm beam}$ is the synthesized beam size in arcsec,
and $D$ = 10.2 kpc. The masses of blue and red lobes are 86 and 70
${\rm M_{\odot}}$, respectively. The dynamical timescale is $t$ =
$3\times10^4$ yr, and the total mass loss rate is $\dot{M}_{\rm
loss}$ = $5.2\times10^{-3}$ ${\rm M_{\sun}~yr^{-1}}$ from the
equations in \citet{gold1984} and \citet{qins2008}. The mass loss
rate $\dot{M}_{\rm loss}$ is less than the mass infall rate
$\dot{M}_{\rm in}$, so the infall is predominant, suggesting the
envelope mass of central star is still increasing rapidly.

From the PV diagram in Figure \ref{fig_pv}, the outflow velocity with 
$V_{\rm out}\approx10.0\,\kms$ was estimated relative to the systemic velocity $V_{sys}\approx52.5\,\kms$.
The momentum of the outflowing gas per year is $\sim0.052$ M$_{\sun}\,\kms$ based on $\dot{M}_{\rm
loss}$ = $5.2\times10^{-3}$ ${\rm M_{\sun}~yr^{-1}}$.
This value is very consistent with the momentum ($\sim0.050$ M$_{\sun}\,\kms$) of the infalling gas per year.
The momentum is conserved in the dynamical process of the HC \HII region. The outflow movement 
can transfer the angular momentum of infalling materials, so that the mass of the central dense core will keep increasing.

\subsubsection{Rotation}

The rotation axis is expected to be along the direction of the outflow, NE-SW.
In Figure \ref{fig_nh3_mom1} for the intensity-weighted mean velocity (moment-1) maps of
NH$_{3}$ (2, 2), there is also an obvious velocity gradient along
the NW-SE direction in a small area, which is evidence for
rotation. However, the velocity gradient direction is inverted for
NH$_{3}$ (3, 3), possibly due to higher noise. We tried to find other
evidence of rotation with the position-velocity diagram method of NH$_{3}$,
but failed. Higher resolution observations are necessary to
unveil whether and how the rotation is precessing in HC \HII region G35.58-0.03.

\subsection{Ammonia Absorption Lines}
\label{sect:ammonia}

The metastable NH$_3$ (2, 2) and (3, 3) arise from energy levels of
65 K and 125 K, respectively; they usually trace very compact gas
${n\rm(H_{2})\sim10^5 - 10^6~cm^{-3}}$\citep{ho1983,chur2002}.
Figure \ref{fig_nh3_spec} presents the molecular spectral lines of
NH$_3$ (2, 2) and (3, 3) with $V_{sys}=52.5~\kms$ at the position of
the 1.3 cm continuum peak. 
The velocity-integrated contours
(Figure \ref{fig_nh3_mom0}) of NH$_3$ (2, 2) and (3, 3) present a
dense structure in front of the continuum with low velocity dispersion (Figure
\ref{fig_nh3_mom2}). The kinematic of the NH$_3$ (2, 2) and (3, 3) lines
is less obvious than that for the HC \HII region G20.08-0.14N
\citep{galv2009}, which shows evidence for accretion. The satellite
hyperfine lines of NH$_3$ (2, 2) and (3, 3) are very noisy. We just
detect the ammonia absorption lines, but no emission lines.

To calculate the optical depth, excitation temperature, and column density of the
molecular gas, we assume that the beam filling factors and the excitation 
temperatures are equal between the NH$_3$ (2, 2) and (3, 3) main line, and in LTE.
Figure \ref{fig_nh3_mom0} shows that the 1.3 cm continuum background almost covers the molecular gas 
distribution, so the apparent optical depth can be derived from the absorption depth \citep{keto1987}
     \begin{equation}
      \tau_{\rm app}=-ln\left(1+\frac{S_{\rm NH_3}}{S_{\rm 1.3\,cm}}\right),
     \end{equation}
where $S_{\rm NH_3}$ and $S_{\rm 1.3\,cm}$ are estimated from the peak flux density of line and continuum.
The apparent optical depths of NH$_{3}$ main line are obtained with $\tau_{\rm app}(2, 2)=0.65$ and $\tau_{\rm app}(3, 3)=0.47$.

Assuming $h\nu\ll kT_{\rm ex}$ and $\tau(2, 2)=\tau_{\rm app}(2, 2)=0.65$, the excitation temperature 
can be derived from equation (A17) in \cite{mang1992} with
\begin{eqnarray}
    T_{\rm ex}(3,3;2,2)  =  -59.1{\rm K}\{ln[\frac{0.847\Delta V(3,3)}{\tau(2, 2)\Delta V(2,2)} 
   \times \nonumber \\       
      ln(1-\frac{S_{\rm NH_3}(3,3)}{S_{\rm NH_3}(2,2)}(1-e^{-\tau(2,2)}))]\}^{-1}, 
\end{eqnarray}
where the $\Delta V$ is the velocity width in $\kms$, and the $S_{\rm NH_3}$ is the line flux density.
The excitation temperature of the NH$_{3}$ main line is $T_{\rm ex}$(3, 3; 2, 2) = 63 $\pm$ 5 K for the peak position. Then,
the column density of $N_{\rm NH_3}(2, 2)$ \citep{mang1992} is 
     \begin{eqnarray}
      N_{\rm NH_3}(2, 2) &=& 3.11 \times 10^{14} {\rm cm^{-2}}
               \frac{T_{\rm ex}(2,2)}{\nu(2,2)}\tau(2,2)\Delta V,
                   \label{eq:nh3_n}
     \end{eqnarray}
where the $\nu(2,2)$ is the line frequency in GHz, and $\Delta V$ is the velocity width in $\kms$.
The resulting column density is $N_{\rm NH_3}(2, 2)$ = 2.5 $\times\,10^{15}\,\rm cm^{-2}$, and
the abundance ratio between NH$_3$(2, 2) and H$_2$ is [NH$_3$(2, 2)]/[H$_2$] = 10$^{-7}$ \citep{ho1983}.

\section{Conclusions}
\label{sect:conclu}

We have reported high angular resolution observations carried out
with the SMA at 1.3 mm and the VLA at 1.3 cm toward the HC \HII region
G35.58-0.03. Combining spectral and continuum data, we have
investigated the dynamical condition and morphological structure of
the star formation region. The main results are summarized as follows.

\begin{enumerate}

\item
With the 1.3 mm SMA and 1.3 cm VLA observations, we detected a total
of 25 transitions of 8 different species and their isotopologues
(CO, CH$_3$CN, SO$_2$, CH$_3$CCH, OCS, CS, H, and NH$_{3}$).
We presented the Gaussian fitting spectra, the moments 0, 1, and 2
maps, and continuum images. The systemic velocity
$V_{sys}\approx52.5~\kms$ was derived from the mean value of the velocities from
Gaussian fits.

\item
G35.58-0.03 is an HC \HII core with electron temperature
$T_e^*=5500$\,K, emission measure EM $\approx
1.9\times10^{9}$\,pc~cm$^{-6}$, local volume electron density $n_e=
3.3\times10^{5}$\,cm$^{-3}$, and broad radio recombination line emission
with FWHM $\approx$ 43.2~$\kms$. The intrinsic core size is
$\sim$3714 AU. We distinguish the free-free emission (25\% $\sim$ 55\%) from the warm dust 
component (75\% $\sim$ 45\%) at 1.3 mm continuum, from the continuum SED fitting among
3.6 cm, 2.0 cm, 1.3 cm, 1.3 mm, 0.85 mm, and 0.45 mm. An early-type star equivalent to
an O6.5 star is postulated to have formed within the HC \HII region based on the derived
Lyman continuum photon number.

\item
Both the H30$\alpha$ and H38$\beta$ lines have nearly the same FWHM $\approx$ 43.2~$\kms$. 
The observed line width $\Delta V$ is due to the dynamical and/or turbulent movements of hot 
ionized gas with $\Delta V_{dyn}\approx$~40.1\,km~s$^{-1}$, the thermal broadening contaminates 
$\Delta V_{ther}$ $\approx$ 15.87\,km~s$^{-1}$, and the pressure broadening only
$\Delta V_P \approx 0.02\sim0.06$\,km~s$^{-1}$.
In addition, the H30$\alpha$ line, probably participating in dynamical movements, shows 
evidence of blueshifted and redshifted wings. Therefore, H30$\alpha$ shows evidence of 
an ionized outflow driving a molecular outflow.

\item
The molecular envelope shows evidence of infall and outflow
with an infall rate of 0.033 M$_{\sun}$~yr$^{-1}$ and a mass loss rate of
$5.2\times10^{-3}$ ${\rm M_{\sun}~yr^{-1}}$, suggesting that the high 
accretion rate might have quenched the \HII region inside. Both the momentum of the 
infalling and outflowing gas per year are $\sim0.05$ M$_{\sun}\,\kms$. 
A collimated bipolar outflow is detected in the
moment-0 maps of CO (2-1), $^{13}$CO (2-1), $^{13}$CS
(5-4), and SO$_2$ 11(1,11)-10(0,10). The intensity-weighted mean
velocity (moment-1) map of NH$_{3}$ (2, 2) has an obvious velocity
gradient along the NW-SE direction, which is an indication for rotation.

\acknowledgements
We would like to thank the anonymous referee for her/his helpful suggestions and comments.
We thank the SMA and VLA staff for making the observations possible.

\end{enumerate}

\bibliographystyle{apj}
\bibliography{references}







\end{document}